\documentclass[a4paper,fleqn]{cas-dc}

\usepackage[authoryear]{natbib}

\def\Method{DeepBVC\@\xspace}

\def\tsc#1{\csdef{#1}{\textsc{\lowercase{#1}}\xspace}}
\tsc{WGM}
\tsc{QE}

\begin{document}
\let\WriteBookmarks\relax
\def\floatpagepagefraction{1}
\def\textpagefraction{.001}

\shorttitle{}    

\shortauthors{G. Zhan et al.}  

\title [mode = title]{Learning from pseudo-labels: deep networks improve consistency in longitudinal brain volume estimation}  

\tnotemark[<tnote number>] 


%

\author[1,2]{Geng Zhan}






\affiliation[1]{organization={Brain and Mind Center, The University of Sydney},
            city={Camperdown},
            postcode={2050}, 
            state={New South Wales},
            country={Australia}}

\author[1,2]{Dongang Wang}[]





\affiliation[2]{organization={Sydney Neuroimaging Analysis Center},
            city={Camperdown},
            postcode={2050}, 
            state={New South Wales},
            country={Australia}}

\author[1]{Mariano Cabezas}[]
            
\author[4]{Lei Bai}[]

\author[1, 2]{Kain Kyle}

\author[4]{Wanli Ouyang}[]

\affiliation[3]{organization={School of Electrical and Information Engineering, The University of Sydney},
            city={Camperdown},
            postcode={2050}, 
            state={New South Wales},
            country={Australia}}
            
\author[1,2]{Michael Barnett}[]

\author[1,2]{Chenyu Wang}[]

\affiliation[4]{organization={Shanghai AI Laboratory},
                city={Xuhui},
                postcode={200232},
                state={Shanghai},
                country={China}}

\cormark[1]

\ead{chenyu.wang@sydney.edu.au}

\cortext[1]{Corresponding author. }



\begin{abstract}
Brain atrophy is an important biomarker for monitoring neurodegeneration and disease progression in conditions such as multiple sclerosis (MS). An accurate and robust quantitative measurement of brain volume change is paramount for translational research and clinical applications. This paper presents a deep learning based method, DeepBVC, for longitudinal brain volume change measurement using 3D T1-weighted MRI scans. Trained with the intermediate outputs from SIENA, DeepBVC is designed to take into account the variance caused by different scanners and acquisition protocols. Compared with SIENA, DeepBVC demonstrates higher consistency in terms of volume change estimation across multiple time points in MS subjects; and greater stability and superior performance in scan-rescan experiments. Moreover, the results also show that DeepBVC is insensitive to acquisition variance in terms of imaging contrast, voxel resolution, random bias field and signal-to-noise ratio. Measurement robustness, automation and processing speed suggest a broad potential of DeepBVC in both research and clinical quantitative neuroimaging applications.
\end{abstract}


\begin{highlights}
\item Our deep learning model is trained with pseudo-labels from SIENA and the impact of label noise is ameliorated by a regularisation method.
\item Data augmentation is introduced to cope with inconsistent imaging acquisitions that are largely unavoidable in clinical settings.
\item Experiments in two datasets demonstrate that \Method has better accuracy, robustness, reliability and consistency compared with SIENA, even though the intermediate outputs from SIENA were used for training.
\end{highlights}


\maketitle

\section{Introduction}\label{intro}
Brain atrophy is a clinically relevant biomarker of disease progression in patients with multiple sclerosis (MS) that reflects irreversible tissue damage due to neuro-axonal destruction, demyelination and gliosis~\citep{bermel2006measurement}.
Accelerated brain tissue loss can be detected in MS cohorts compared to a healthy control population~\citep{de2016establishing}. 
Clinically, brain atrophy is a key predictor of future disease worsening and cognitive impairment in patients with MS~\citep{de2014clinical, jacobsen2014brain, popescu2013brain}
; and has been used frequently in MS clinical trials as a secondary measure of treatment efficacy ~\citep{filippi2004interferon, cadavid2017safety}.
However, the incorporation of brain atrophy into monitoring paradigms for individual patients requires significantly improved accuracy, precision and robustness. Several algorithms~\citep{hajnal1995detection, friston2003statistical, bermel2003semiautomated, horsfield2003whole, rudick1999use, collins2001automated} have been developed for quantifying longitudinal brain volume change (BVC).
The input for these algorithms is a pair of T1-weighted (or FLAIR) images at baseline and follow-up time points.
For most methods, it is common practice to perform linear registration to align the baseline and follow-up images, and brain segmentation, before volume change analysis. Boundary Shift Integral~(BSI)~\citep{freeborough1997boundary}, gBSI~\citep{prados2015measuring}, and SIENA~\citep{smith2002accurate} use different measurements to track the movement of the brain boundary.
IPCA~\citep{chen2004automated} uses an iterative principal component analysis method to find the outliers reflecting between-scan differences.
MSmetrix~\citep{smeets2016reliable} ) adopts nonrigid registration and Jacobian integration of deformation fields to produce atrophy measures.
Similarly, Quantitative Anatomical Regional Change~(QUARC)~\citep{holland2011nonlinear} utilises non-rigid registration, but directly calculates hexahedral volumes to yield the fractional volume change. 
FreeSurfer-longitudinal~(FS)~\citep{reuter2012within} is a segmentation-based method that performs tissue segmentation at each time point. 
Among those methods, SIENA~\citep{smith2002accurate} is arguably the most widely used algorithm in MS clinical trials. 

Although continuous efforts have been made to develop new methods that improve the accuracy and reliability of BVC quantification, longitudinal MRI measurement is susceptible to inconsistency of imaging acquisition at baseline and follow-up~\citep{lee2019estimating, medawar2021estimating}, ), particularly in routine clinical practice. 
Common examples of inconsistencies that impact the reliability of quantitative BVC measurement, regardless of methodology, include image contrast differences ~\citep{preboske2006common}, intensity non-uniformity~\citep{takao2010effects}, noise, or different resolutions and voxel spacing~\citep{vrenken2013recommendations}).
Additionally, there are few comparative studies that assess the performance of newer (versus older) measurement methods, such as SIENA and Boundary Shift Integral~\citep{freeborough1997boundary} respectively, in the presence of such acquisition inconsistencies.

However, several approaches have been proposed to ameliorate the influence of acquisition inconsistencies and thereby improve the accuracy of BVC measurement algorithms.

The first approach aims to reduce scan inhomogeneity during pre-processing~\citep{lewis2004correction, smith2002accurate, vemuri2005coil, vovk2004mri, learned2004joint, duffy2018retrospective, higaki2019improvement}.
By removing the bias field from the longitudinal input scans, these methods aim to reduce the variance in BVC estimates. 
Several other data harmonisation methods~\citep{liu2021style, dewey2019deepharmony, dewey2020disentangled, beer2020longitudinal, garcia2020neuroharmony} aim to improve the qualitative and quantitative consistency of differently acquired MRI scans. 
In practice, these methods assign one scan as a reference and process images in the second scan to narrow the differences  attributable to protocol inconsistency. Although many of these methods aim to improve the quantitative utility of MRI in long-term or multi-site studies, most are not specifically designed for consistent BVC measurement.  Rather, they focus on harmonisation of images to a reference image or providing segmentation masks with greater consistency (as determined by Dice, Coefficient of Joint Variation or similar) with those derived from a reference image in a test-retest setting. Therefore, the effectiveness of these methods for producing consistent BVC measurements is not directly measured.

The second approach focuses on correcting BVC estimates during post-processing. 
\cite{lee2019estimating} estimates the fixed effect of scanner changes with a linear model and accounts for this factor during measurement of BVC. 
\cite{sinnecker2022brain} estimates an additive fixed corrective term for scanner change by comparing BVC rates during scanner change and no scanner change for healthy control subjects. These methods are subject to scanner-specific variation and require group level test results to estimate the correction factor.

The addition of both pre-processing and post-processing steps to BVC measurement algorithms increase overall computational complexity and processing time. Additionally, while these methods may result in qualitative and quantitative improvements, most are confined to the research domain and their integration with (and effectiveness in) clinical workflows is not clear.


In this paper, we introduce DeepBVC, a novel framework that combines a deep neural network with data augmentation to provide fully automated and robust BVC assessment.  
A deep neural network offers generalisation to common brain atrophy patterns; and comprehensive data augmentation~\citep{shorten2019survey} provides robustness and mitigates protocol and other acquisition-related inconsistencies. 
Specifically, the deep neural network module is used to estimate shift at the brain boundary from baseline to follow-up; and the data augmentation module synthesises images that contain common image distortions and acquisition differences during the training stage of the deep neural network. We also propose a novel training regime for DeepBVC. In general, supervised neural network training requires a large-scale dataset with accurate labels. However, it is not practical to acquire accurate sub-voxel level atrophy estimations from images.
Existing BVC measurement algorithms contain known or unknown bias and random variation factors for different scans~\citep{thanellas2010sensitivity}.
Therefore, we used the brain boundary shift produced by SIENA as the pseudo-label in training, noting precedents for the use of software-generated annotations as the label for tasks such as brain parcellation~\citep{henschel2020fastsurfer}.
We then utilised the inherent regularisation properties of convolutional neural networks to tackle the underlying label noise ~\citep{goodfellow2014explaining, zhang2017regularizing, zhang2017mixup}.

In summary, \Method is a deep learning based method  for longitudinal BVC assessment with the following contributions:
\begin{itemize}
    \item Data augmentation is introduced to cope with inconsistent imaging acquisitions that are largely unavoidable in clinical settings.
    \item Our deep learning model is trained with pseudo-labels from SIENA and the impact of label noise ameliorated by a regularisation method.
    \item Experiments in two datasets demonstrate that \Method has better accuracy, robustness, reliability and consistency compared with SIENA, even though the intermediate outputs from SIENA were used for training.
\end{itemize}

\section{Materials and Methods}
This section is organised as follows: 
We describe the data used in this study in Sec.~\ref{sec:materials} and the data preprocessing in Sec.~\ref{sec:preprocessing}.
In Sec.~\ref{sec:methods}, we illustrate the details of our method, including the model structure, training details, and the integration our model into the pipeline of BVC estimation.
Next, we introduce the settings and metrics of evaluation experiments in Sec.~\ref{sec:experimenta_design}.

\subsection{Data Acquisition}
\label{sec:materials}
We use two data sources for this study: an in-house dataset (MS Clinical Dataset) comprising clinical data and matched MRI scans from patients with relapsing remitting multiple sclerosis (RRMS) subjects; and a public test-retest dataset (Maclaren test-retest dataset)~\citep{maclaren2014reliability} comprising scans from three healthy control subjects.
The study was approved by the University of Sydney and followed the tenets of the Declaration of Helsinki.

\begin{figure*}[]
	\centering
		\includegraphics[scale=0.5]{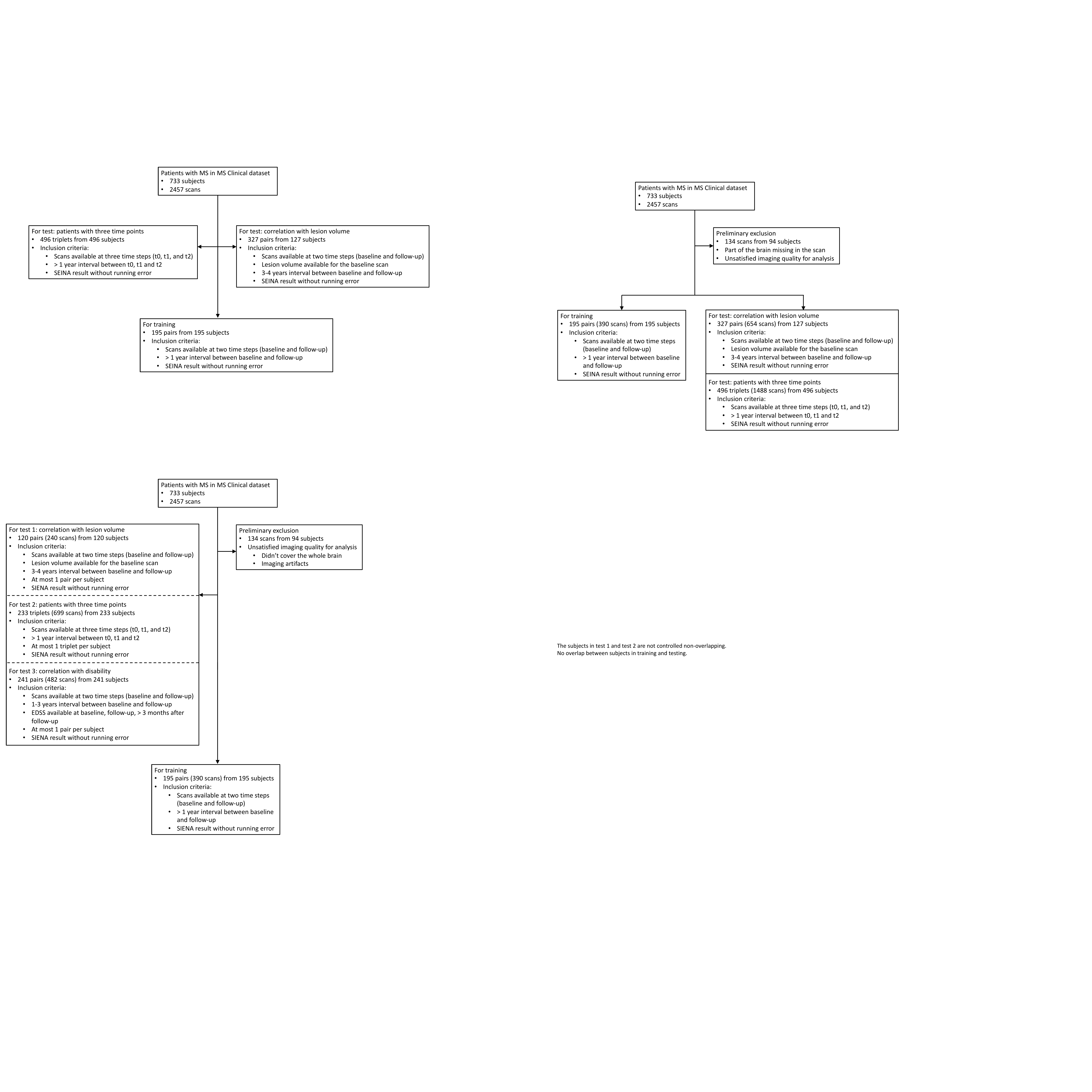}
  \vspace{10pt}
	  \caption{
        Baseline data for the MS Clinical Dataset. Non-overlapping constraints were applied between the subjects included in training and testing, but overlapping was permitted for data involved in the three testing experiments.}\label{fig:patient_disposition}
\end{figure*}

\begin{table*}[<options>]
\caption{Demographic, clinical characteristics of the patients in the in-house dataset  at baseline.} \label{tab:demographic}
\begin{tabular*}{\tblwidth}{LLLLL}
\toprule
 & train & test: lesion correlation & test: three time points & test: disability correlation \\ 
\midrule
 Patients, n (\% female) &  195 (74) & 120 (81) & 233 (77) & 208 (33)\\
 Age, mean (SD) (years) & 41.6 (12.4) & 40.3 (10.5) & 41.5 (9.6) & 39.7 (9.4) \\
 Disease Duration, mean (SD) (years) & 9.4 (8.7) & 8.7 (7.9) & 10.4 (6.9) & 8.5 (6.0) \\
 EDSS\footnotemark{}, mean (SD) & 2.1 (1.8) & 1.9 (1.6) & 1.6 (1.8) & 2.0 (1.8) \\
\bottomrule
\end{tabular*}
\end{table*}

\footnotetext{Expanded Disability Status Scale}

\subsubsection{MS Clinical Dataset}
\label{sec:ms_clinical}
Written informed consent was obtained from all participants. In total, 2457 T1-weighted MRI exams from 648 patients diagnosed with RRMS were included in this study (Tab.~\ref{tab:demographic}.
All patients were recruited from the MS Clinic based at the Brain and Mind Centre, University of Sydney; and clinical MRI exams were acquired with one of three 3T MRI scanners (Tab.~\ref{tab:acquisition}) between 2010 and 2020. 
Longitudinal scans for each patient were acquired with the same scanner using a consistent protocol. 
MRI acquisition parameters are summarised in Tab.~\ref{tab:acquisition}.


\begin{table}[]
\caption{MRI acquisition details for the MS Clinical Dataset.}
\label{tab:acquisition}
\begin{tabular}{lccc}
\hline
Scanner & T1 parameters \\
\hline
GE Discovery 3.0T & \begin{tabular}{c}TE = 2.6ms \\ TR = 7ms \\ TI = 0.45s \\ $0.93 \times 0.93 \times 1 mm^3$ \end{tabular} \\
\hline
Philips Ingenia 3.0T & \begin{tabular}{c} TE = 2.4ms \\ TR = 8ms \\ $1 \times 1 \times 1 mm^3$ \end{tabular} \\
\hline
SIEMENS Skyra 3.0T & \begin{tabular}{c} TE = 2.5ms \\ TR = 2.2s \\ TI = 0.9s \\ $0.90 \times 0.90 \times 0.90 mm^3$ \end{tabular} \\
\hline
\end{tabular}
\end{table}

\subsubsection{Test-retest Data}
\label{Sec:test-retest}
We use the Maclaren test-retest dataset \citep{maclaren2014reliability} to test measurement reliability. 
The dataset comprises 120 T1-weighted scans from 3 healthy subjects  aged 26-31 years, acquired with a GE MR750 3T scanner. Each subject was scanned twice on 20 different days within a 31-day period. The acquisition protocol (TE: 3ms, TI: 0.4s, TR: 7.3ms, 1.2mm slice thickness) followed the recommendations of the Alzheimer’s Disease Neuroimaging Initiative (ADNI)~\citep{jack2008alzheimer}.



\begin{figure*}[<options>]
	\centering
		\includegraphics[scale=0.7]{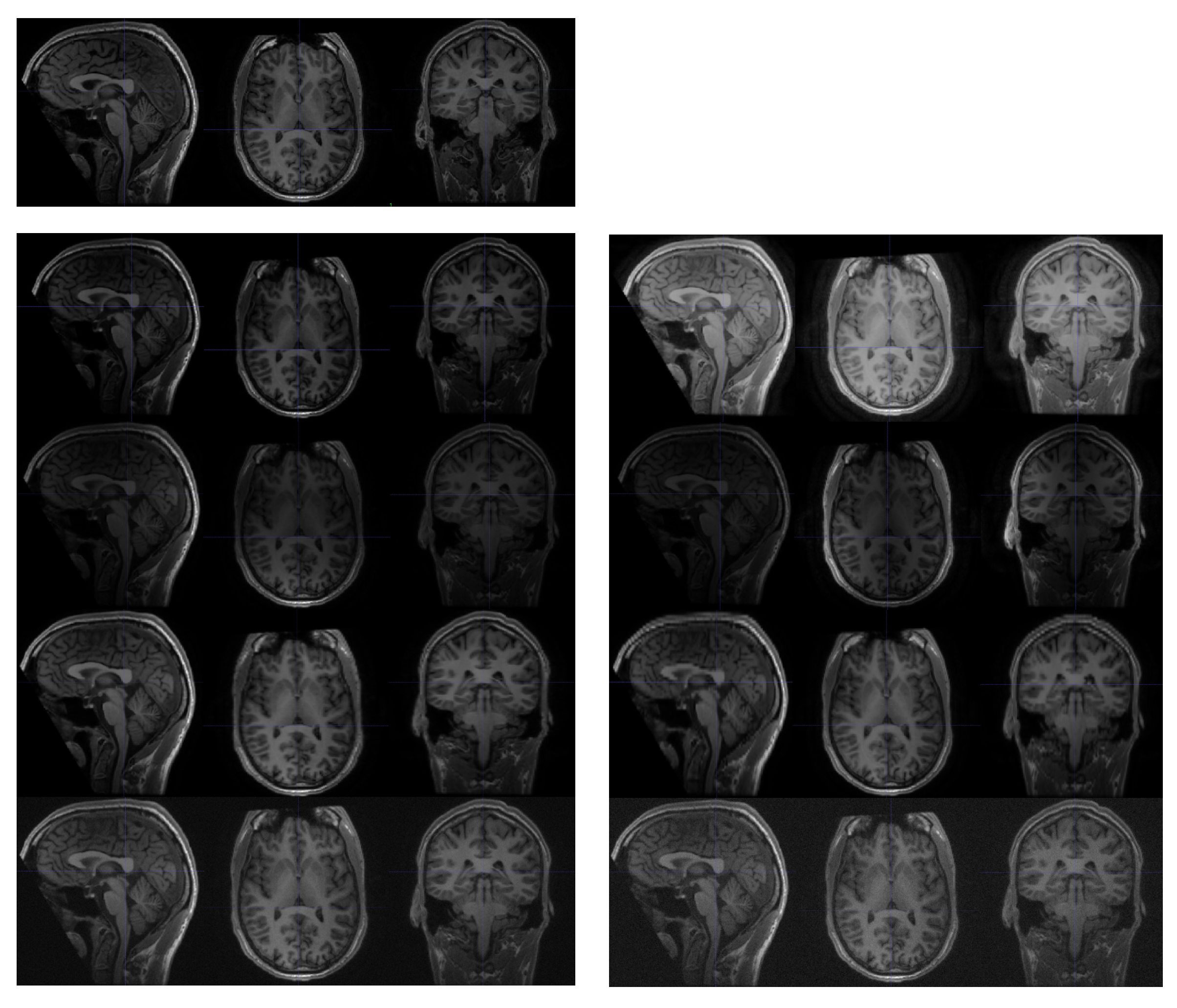}
            \label{fig:synthesised_protocol_inconsistency}
	  \caption{
 Examples of synthesised scans for the test-retest dataset. All scans are shown in sagittal, axial, and coronal planes. The upper-left images are slices from the original scan; and the following two columns are synthesised from the original scan. Each row (from top to bottom) represents a different acquisition artifact: contrast, bias field, spacing anisotropy, and Gaussian noise. The images on the left are less distorted than those shown on the right.}\label{}
\end{figure*}

\subsection{Data Pre-processing}
\label{sec:preprocessing}

\subsubsection{Format Conversion and Data Selection}
For the MS Clinical Dataset, images were originally stored in DICOM format. All acquired DICOM data were converted to NIFTI format by $dcm2nii$~\citep{li2016first}. The quality of all MRI data was visually assessed by experienced neuroimaging analysts at the Sydney Neuroimaging Analysis Centre (Sydney, Australia). Images that failed quality assessment (incomplete brain coverage, severe imaging artifacts) were excluded from further study.  N4 Bias Field Correction~\citep{lowekamp2013design} was applied to remove the bias field from all scans meeting quality criteria.

\subsubsection{Brain and Skull Segmentation}
Brain and skull segmentation was performed with BET \citep{jenkinson2005bet2} for all eligible scans and the masks were manually refined by experienced neuroimaging analysts. Skull stripping was undertaken to generate images only containing brain tissues for subsequent analysis.


\subsubsection{SIENA Analysis}

\textbf{Co-registration:} 
For the longitudinal scan pairs of each subject, T1-weighted scans at baseline and follow-up were aligned using two-step affine registration~\citep{freeborough1997boundary, smith2002accurate}. First, the skull masks were used to optimise the scale and skew; then, the brain images were used to optimise image translation and rotation. All registration results were manually checked by trained neuroimaging analysts, and poorly aligned pairs excluded from further analysis.

\textbf{Brain edge point segmentation:} 
After brain alignment, FAST~\citep{zhang2001segmentation} was used to segment longitudinal scans into the principal brain tissue compartments: white matter (WM), grey matter (GM), and cerebrospinal fluid (CSF). As whole brain volume change includes changes for both WM and GM, the probabilistic maps from FAST were binarised (using a threshold of 0.5) and defined the union of WM and GM as the foreground segmentation, and the remainder of the image as background. Consequently, brain edge points were defined by the edges of the foreground mask.

\textbf{Change analysis:}
Voxel-wise atrophy/growth was estimated for all brain edge points from baseline to follow-up time points; and the mean edge point motion converted into PBVC (a single number) with a self-calibration step.



\subsection{Model}
\label{sec:methods}

\begin{figure*}[<options>]
	\centering
		\includegraphics[scale=0.5]{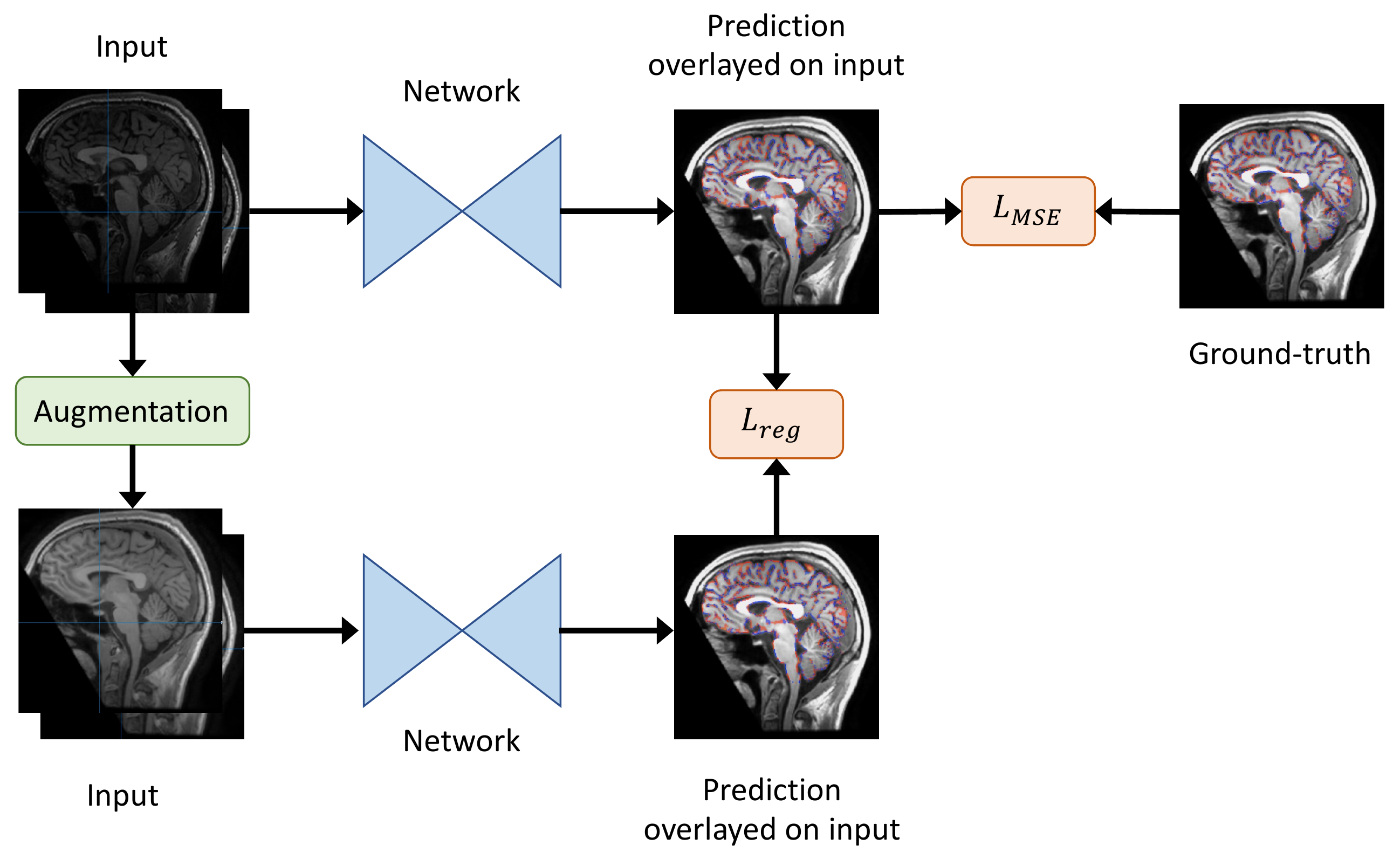}
	  \caption{
        An overview of our method. Given the original inputs (a pair of baseline and follow-up scans, and the mask of baseline brain edge points), new inputs are generated by augmentation (including re-sampling and random contrast) without spatial transformation. As a result, the newly generated inputs share the same labels as the original inputs. Both the generated and original inputs are fed into the same network. Then, the output of the original inputs is compared to the label to obtain the loss $L_{MSE}$. Similarly, the output of the original inputs is compared to that of the generated inputs to yield the regularization term $L_{reg}$. Finally, the total loss is defined as in Eqn.~\ref{eqn:loss}.
   }\label{fig1}
\end{figure*}

\subsubsection{Model Structure}
\label{sec:network}
We used a 3D-Unet~\citep{cciccek20163d} as the backbone network for feature extraction, followed by a single convolutional layer as the prediction head. The main blocks of the networks use residual convolutional layers ~\citep{he2016deep} with group normalisation~\citep{wu2018group}.
The model inputs comprise a pair of baseline and follow-up T1 images. The label is a voxel-wise estimation of brain boundary shift produced from SIENA. Because SIENA estimates the brain boundary shift for each voxel location in the input, the label and the model output are 3-D tensors with the same dimensions as the inputs. During both training and inference, the model output is masked with the brain boundary segmentation and only the outputs at the boundary locations are retained (outputs are set to 0 for non-boundary locations).

\subsubsection{Training Details}
For each iteration, 8 scan pairs were randomly selected from all training scan pairs. Each pair was randomly cropped into a pair of patches of size $64 \times 64 \times 64$ due to GPU memory constraints. The patches were then fed into the model and the loss calculated accordingly to update the model weights.

We used mean squared error (MSE) as the loss function $L_{MSE}$ to evaluate the model's deformation prediction when compared with the pseudo-labels from SIENA (edge point motion). 
Furthermore, to render the model insensitive to differences in imaging quality during training, we adopted a consistency regularisation loss. Specifically, the loss minimised the difference between the predictions derived from the original images and the augmented images, such that brain volume differences were maintained in the context of an isolated change in imaging acquisition conditions. 

Formally, for a data point $x$, the regularisation loss term was defined as:

\begin{equation}
\label{Eqn:1}
    L_{reg} = \lVert p_{model} ( y |x; \theta) - p_{model} (y | Augment(x); \theta) \rVert ^2,
\end{equation}

where $Augment(x)$ is a stochastic transformation, whose effect is not identical for each training sample, but rather follows a distribution. As the regularisation term requires the underlying brain volume change to remain constant, spatial transformations that deformed the original brains were not permitted. Therefore, we included both random spacing anisotropy re-sampling and random contrast alterations as possible augmentation steps. For clarity, only one type of augmentation was randomly selected for each sample.

Finally, the overall loss $L$ was defined as:

\begin{equation}
    L = L_{MSE} + \lambda L_{reg},     
    \label{eqn:loss}
\end{equation}

where we set $\lambda = 0.01$. 

The final loss function was optimised using the Adam optimiser~\citep{kingma2014adam} with an initial learning rate of 0.001 that was reduced by a factor of 0.5 when the loss did not drop for 3 consecutive epochs. The model was trained for 2500 iterations per epoch and for 50 epochs in total. Model optimisation was performed with 2 NVIDIA V100 GPU cards on an NVIDIA DGX-1 station.

\subsubsection{Training Data}
\label{sec:training_data}
To train the model, we collected 195 pairs of scans (1 pair per subject) from the MS Clinical Dataset (Sec.~\ref{sec:ms_clinical}, Fig.~\ref{fig:patient_disposition}), from which 70\% (137 pairs) were randomly selected for training and the remaining 30\% (58 pairs) for validation. Two additional, independent imaging datasets were used for testing as described in Tab.~\ref{tab:demographic}. Testing datasets did not overlap with training or validation datasets; and all 195 subjects involved in model development were excluded from evaluation experiments.

\subsection{Experimental Setup}
\label{sec:experimenta_design}
Five experiments were undertaken to evaluate the performance of DeepBVC with respect to test-retest consistency, multi-step longitudinal consistency, protocol-agnostic test-retest consistency, relevance to T2 lesion and correlation with disability.

\subsubsection{Consistency with Test-retest Data}
\label{design:test-retest}
The Maclaren test-retest data used for this experiment is described in Sec.~\ref{Sec:test-retest}. We grouped the data into baseline follow-up pairs as follows: the two scans from the same day and subject were used as a longitudinal pair with no atrophy. We used 60 pairs in total (20 pairs per subject).

For this experiment, we assumed that there would be no atrophy because the pairs were scanned back-to-back (i.e. the brain volume difference should be 0\%). We ran SIENA and DeepBVC methods to measure the brain volume difference for each pair. We then grouped the results by subject and report each method’s mean and standard deviation. We also report the mean absolute error for the results of each subject, where the error was acquired by comparing the BVC measurements against the 0\% BVC.

\subsubsection{Influence of the Protocol Inconsistency}
\label{design:protocol_inconsistency}
To test the influence of various acquisition protocol inconsistencies, we used the Maclaren test-retest data (Sec.~\ref{Sec:test-retest}), coupled with imaging processing techniques to synthesise new image pairs with protocol pseudo-inconsistencies, as described below and shown in Fig.~\ref{fig:synthesised_protocol_inconsistency}. We then compared the experimental results with those derived from the original test-retest data. Experiments followed the design described in Sec.~\ref{design:test-retest}, with the addition of a pre-processing step to include synthesised images as follows:

\textbf{Random contrast adjustments.}
Image intensities were adjusted by the parameter $\gamma$.
Each voxel intensity $x$ was updated as 

\begin{equation}
    x = (\dfrac{x - x_{min}}{x_{range}}) ^ \gamma \cdot x_{range} + x_{min},
    \label{eq:contrast}
\end{equation}

where $x_{min}$ is the minimal voxel value in the original brain image, and $x_{range}$ is the difference between the maximal and minimal voxel value in the scan, excluding the background.

\textbf{Random bias field.}
The bias field was generated from a linear combination of smoothly varying basis functions, as described in \cite{van1999automated}. In practice, we observed that intensity inhomogeneity was more likely to occur along one of three axes (sagittal, coronal, and axial). Therefore, we synthesised the random bias field along one of the (randomly selected) axes.

\textbf{Random spatial anisotropy.}
Spacing anisotropy was introduced by down sampling an image along an axis and then re-sampling back to its original spacing. In our experiment, we simultaneously performed this random transformation along all three axes (sagittal, coronal, and axial).

\textbf{Gaussian noise.}
The noise for each voxel was sampled from a normal distribution with random parameters, and added to the original image.

To systematically explore the impact of protocol inconsistency, we analysed the performance of both DeepBVC and SIENA for different levels of the four types of inconsistency, controlled by different parameters during synthesis. For protocol inconsistency in Gaussian noise, spatial resolution anisotropy and bias field, the value of the parameters is positively correlated to the degree of inconsistency. For the inconsistency in image contrast, the measurement follows a different relationship (Eq.~\ref{eq:contrast}), namely image contrast is controlled by $\gamma$. The larger the difference between $\gamma$ and $1$, the higher the inconsistency between synthesised and original images.


\subsubsection{Multi-step Consistency with Three Time Points}
\label{sec:triplet_data}
Inspired by the experiments of \cite{smith2002accurate}, we included data from three time points in our analysis, enabling both single-step and a multi-step measurement strategies. For the single step approach, we estimated brain atrophy between the first (t0) and last (t2) time points, while for the multi-step approach, we combined the estimated intermediate atrophy between the t0 and the second timepoint (t1), and t1 and t2. A smaller difference between the single-step and multi-step approaches indicates a more consistent measurement.

For these experiments, we use data selected from the MS Clinical Dataset (Sec.~\ref{sec:ms_clinical}), restricting inclusion to those subjects with 3 available brain scans with an interval of at least 1 year between successive time points. In total, 233 subjects (3 scans each subject) were used for this experiment.

\subsubsection{Brain T2 Lesion Volume and Brain Volume Change}
\label{sec:corr_lesion}
To further investigate the predictions of our method and their clinical impact, we analysed the correlation between brain volume change and baseline lesion volume as the rate of brain volume loss has been found to correlate with baseline T2 lesion volume~\citep{tedeschi2005brain}. We assumed that improved accuracy of BVC measurement would enhance this correlation. For this experiment, scans were selected from the MS Clinical Dataset (Sec.~\ref{sec:ms_clinical}), based on availability of T1-w and FLAIR sequences at baseline and T1-w images at follow up (3-4 years subsequent to the baseline time point). The baseline lesion volume was derived from an in-house deep learning lesion segmentation algorithm~\citep{liu2022dams} followed by manual review (and, if required, correction) of lesion masks by trained neuroimaging analysts at the Sydney Neuroimaging Analysis Centre. We report linear correlation and partial correlation (controlled for age and disease duration) for DeepBVC and SIENA. SPSS~\citep{field2013discovering} was used for this analysis.

\subsubsection{Disability and Brain Volume Change}
Sustained progression of the expanded disability status scale (EDSS) score has reported to correlate with higher rates of brain atrophy~\citep{rudick2000brain, bermel2006measurement}. To investigate the clinical relevance of the BVC, we therefore compared the annualised BVC with EDSS progression over 1-3 years in subjects with both T1-w scans and clinical data available at two time points with an interval of at least 12 months. Subjects were firstly grouped into EDSS progressors and non-progressors, where EDSS progression was determined by 3 strata as previously described by ~\cite{kalincik2015defining} and confirmed over 3 months. BVC was then determined by both DeepBVC and SIENA and reported for each group.  We also analysed BVC for matched subjects from each group, using 1-to-multiple propensity score matching based on age and disease duration.


\section{Results}
For each validation experiment involving the MS Clinical data, we included all subjects that met the relevant inclusion criteria (Sec.~\ref{sec:methods}, Sec.~\ref{sec:experimenta_design} and Fig.~\ref{fig:patient_disposition}). The demographic and clinical characteristics of the subjects eligible for training and each validation are listed in Tab.~\ref{tab:demographic}. Among 2457 scans from 648 subjects, 134 scans from 94 subjects were first excluded from all experiments due to poor imaging quality. For multi-step consistency experiments (Sec.~\ref{sec:triplet_data}), 233 eligible subjects (77\% female) with 3 available scan timepoints were used.  At the time of baseline imaging, mean age and disease duration was 41.5 and 10.4 years respectively; and mean EDSS was 1.6, in this group. For the lesion experiment, 120 pairs of scans were available. In this group, 81\% of the patients were female, with a mean age and disease duration of 40.3 and 8.7 years respectively; and a mean EDSS of 1.9. The remaining subjects (195 subjects/scan pairs, 74\% female) were used for training. In this group, mean age and disease duration was 41.6 and 9.4 years respectively; and mean EDSS was 2.1.


\subsection{Consistency with Test-retest Data}

\begin{figure*}[<options>]
	\centering
		\includegraphics[scale=0.5]{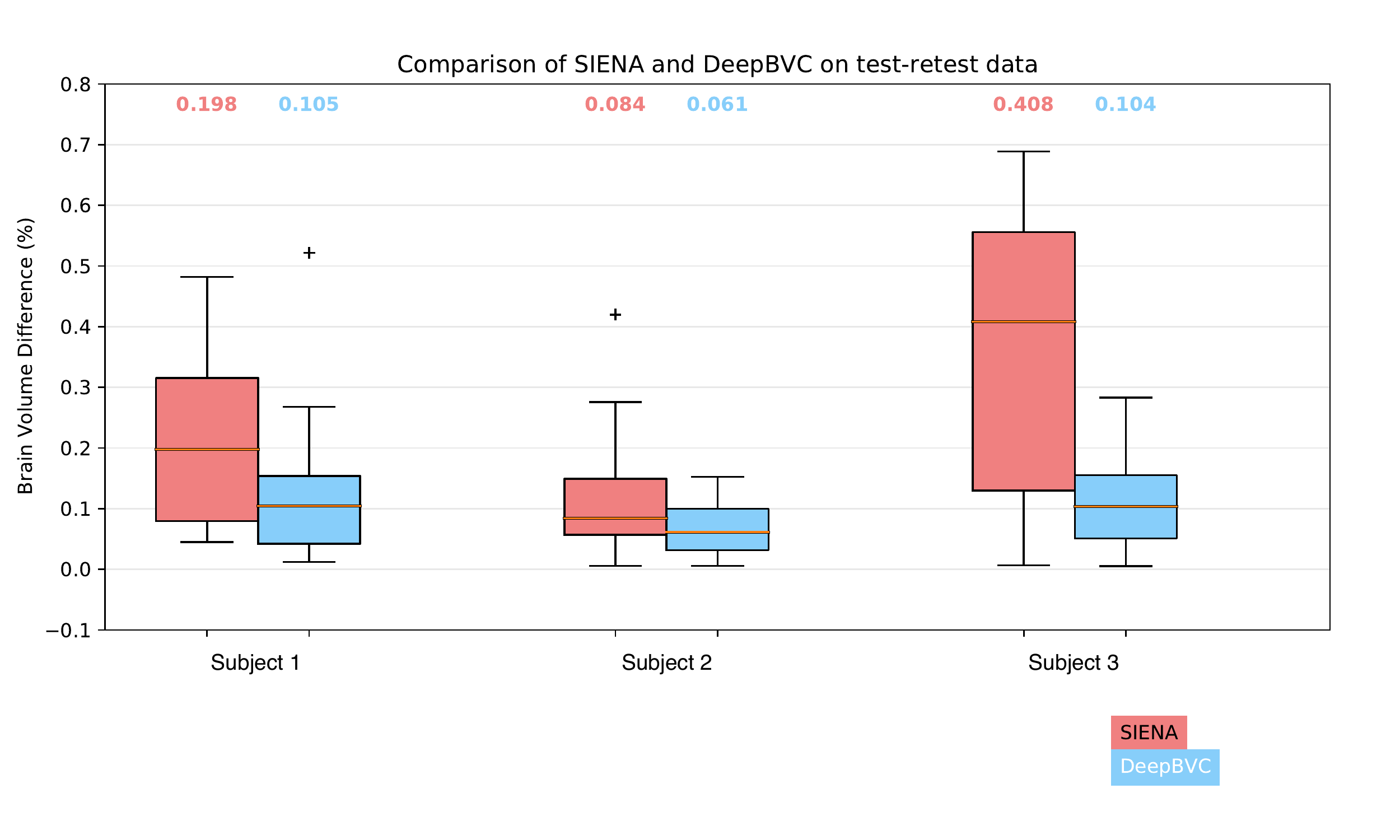}
            \vspace{-15pt}
	  \caption{
   Box plots for SIENA and \Method on test-retest data. Results are grouped and shown for each subject. The median brain volume difference is reported above each corresponding box plot.
   }
   \label{fig:test-retest-original}
\end{figure*}

\begin{table}[<options>]
\caption{
Brain volume difference in percentage for three subjects in the test-retest dataset. The numbers are reported as average ($\pm$ standard deviation) across all the sessions for each subject.
}
\label{tab:original_test_retest}
\begin{tabular*}{\tblwidth}{@{}LLL@{}}
\toprule
 Subject & SIENA (\%) & \Method (\%) \\ 
\midrule
sub1 &  0.212 ($\pm$ 0.139) & 0.126 ($\pm$ 0.116) \\
sub2 & 0.118 ($\pm$ 0.103) & 0.066 ($\pm$ 0.042) \\
sub3 & 0.351 ($\pm$ 0.228) & 0.111 ($\pm$ 0.077) \\
\bottomrule
\end{tabular*}
\end{table}

We illustrate the performance of SIENA and \Method in Fig.~\ref{fig:test-retest-original} and Tab.~\ref{tab:original_test_retest}. The subject-wise means and standard deviations of PBVC measured by DeepBVC were smaller than by SIENA (Tab.~\ref{tab:original_test_retest}). For all three subjects, the BVC measured by DeepBVC was less dispersed and was closer to 0 (Fig.~\ref{fig:test-retest-original}). The median PBVC for DeepBVC smaller than the equivalent plot for SIENA for all subjects (0.105 vs. 0.198, 0.061 vs. 0.084, 0.104 vs. 0.408 respectively). One outlier with a large BVC was found for DeepBVC (subject 1) and another for SIENA (subject 2). 


\subsection{Influence of the Protocol Inconsistency}
\label{sec:experiment_stratified_inconsistency}
The estimated PBVC for both methods was $0$ for pairs of identical scans.

\begin{figure*}[]
	\centering
		\includegraphics[scale=0.04]{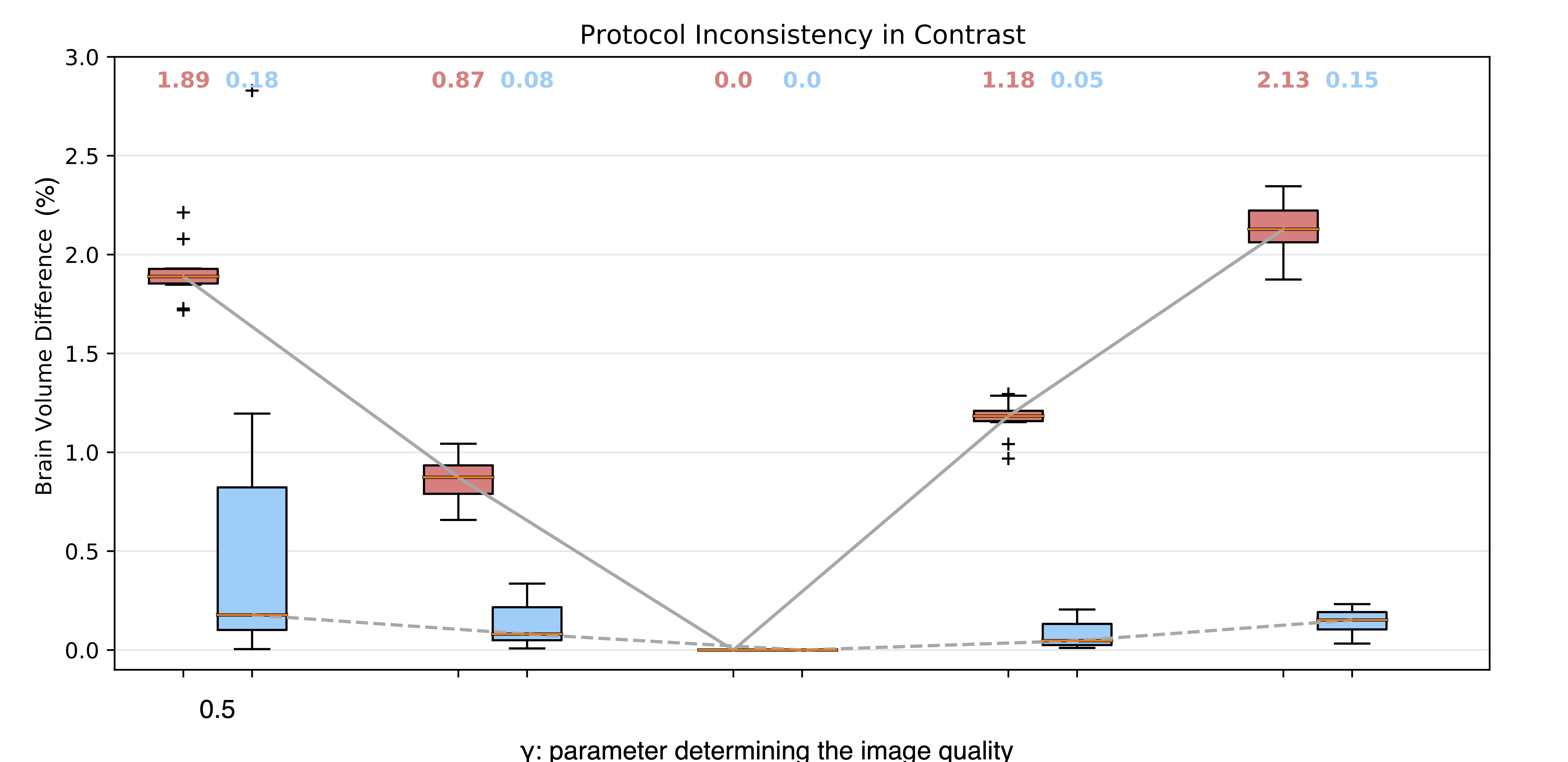}
            \includegraphics[scale=0.04]{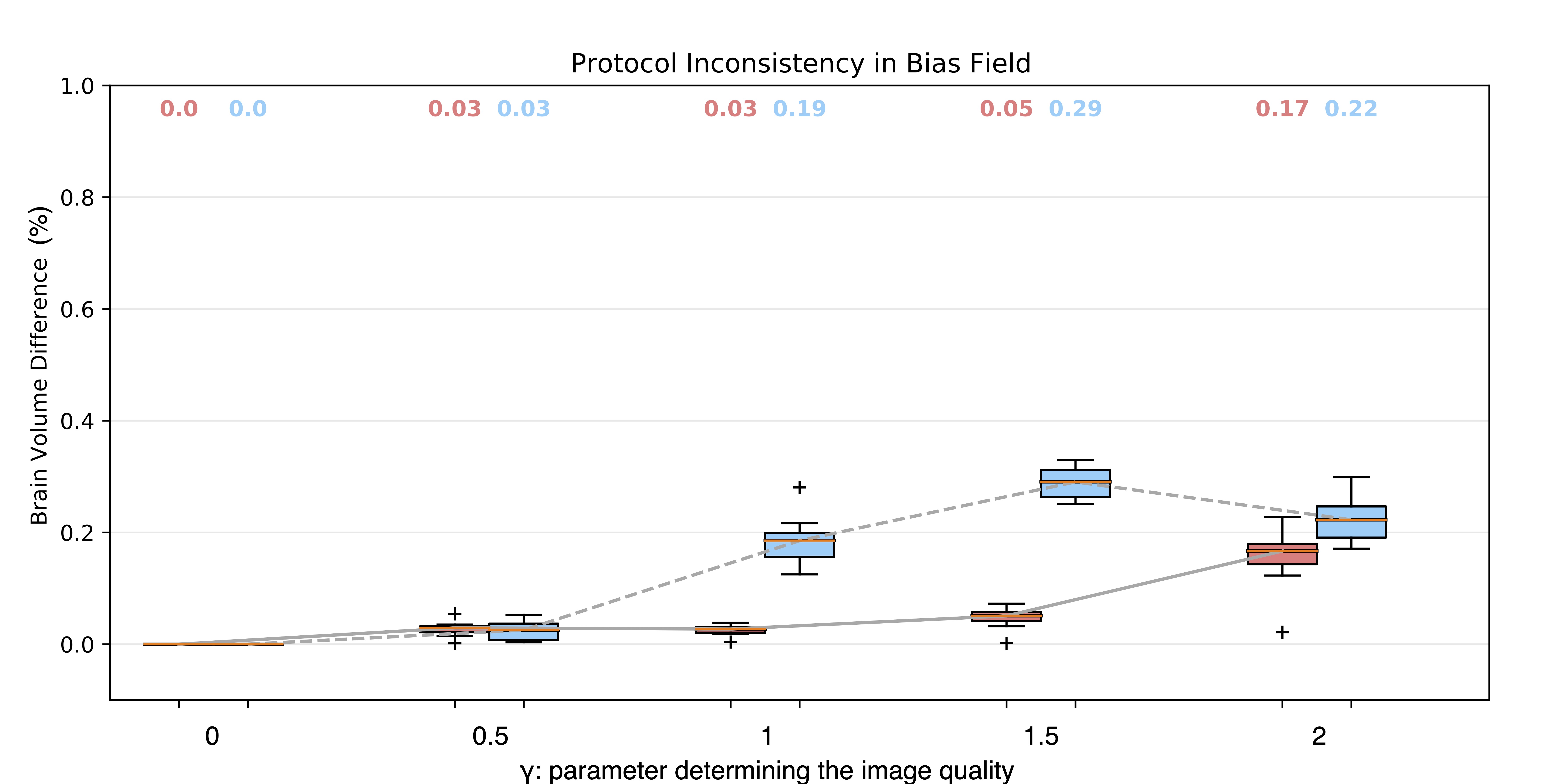}
            \includegraphics[scale=0.04]{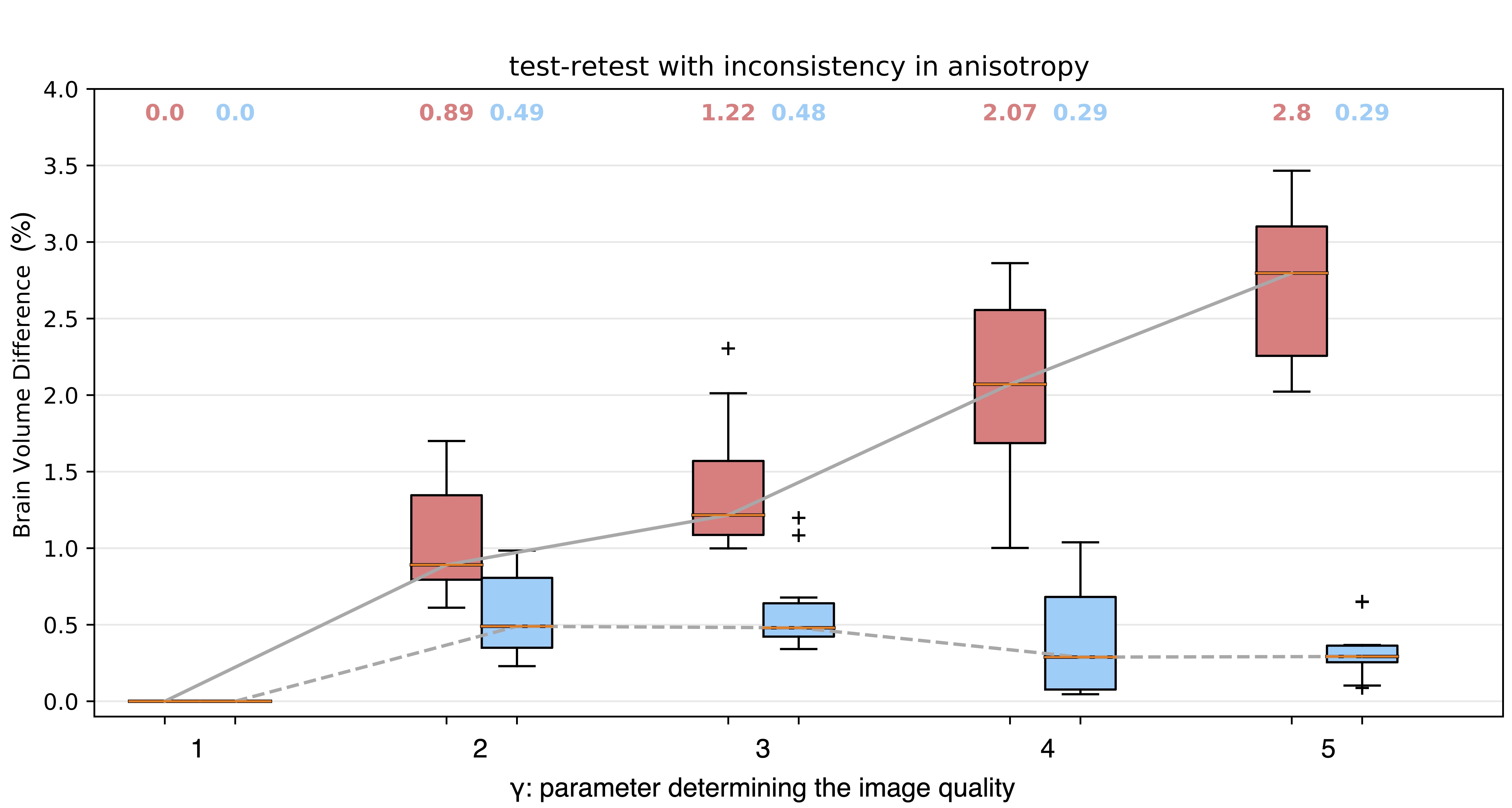}
            \includegraphics[scale=0.04]{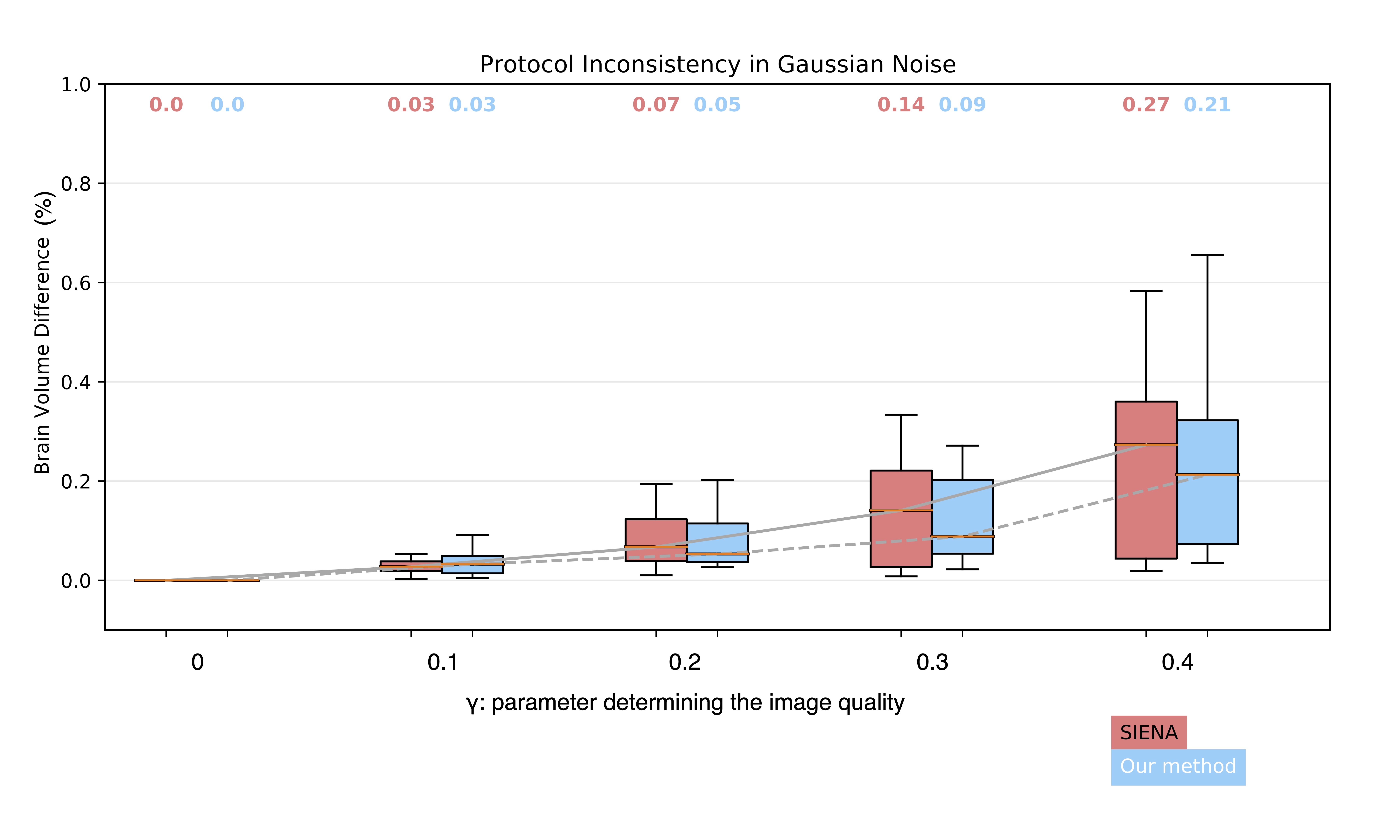}
            \vspace{-10pt}
	  \caption{
   Brain volume difference estimation for stratified protocol inconsistency. For random contrast, the imaging quality is unchanged when the parameter is 1. The larger the difference between the parameter and 1, the lower the image quality and more inconsistent the protocol. The x-values are parameters determining the image processing step and the output image quality. For random Gaussian noise, bias field and anisotropy, parameters are negatively correlated to the quality of the output image (the higher the value, the lower the image quality, the more inconsistent the protocol). The box plots at the left-most of each figure represent the imaging quality is unchanged. 
   }\label{fig:stratified_inconsistency}
\end{figure*}

For contrast inconsistencies (Fig.~\ref{fig:stratified_inconsistency}), the box plot for DeepBVC showed a much lower distribution of errors than SIENA; and, unlike SIENA, no significant increase as gamma changed (up to $\pm0.5$). Furthermore, the median brain volume difference of DeepBVC was lower than SIENA by one order of magnitude. Finally, the variance in the error was similar for both methods when $\lambda$ was $0.75,1.25,1.5$. However, when $\lambda=0.5$, the errors for DeepBVC showed a wider distribution than SIENA. 


For bias field inconsistencies, the medians and interquartile ranges for DeepBVC were higher than SIENA, though brain volume difference was low for both techniques (range $0.00-0.29$ and $0.00-0.17$ respectively). 


For spatial resolution anisotropy, the differences measured by SIENA became greater as the inconsistency level increased. As shown in Fig. 5, the median difference gradually increased from 0.89\% ($\gamma=2$) to 2.8\% ($\gamma=5$). For DeepBVC, the median difference remained low at the different levels of inconsistency tested. For the highest level of inconsistency, the median difference measured by SIENA reached 2.8\%, while DeepBVC remained as low as 0.29\%. The error distribution of DeepBVC was less scattered and closer to 0 than the equivalent of SIENA for all levels of inconsistency tested.


For Gaussian noise, the error distribution of DeepBVC was less dispersed and closer to 0 than the equivalent of SIENA when the Gaussian noise level was 0.2, 0.3 and 0.4. The error distribution of DeepBVC was wider than SIENA and the median was 0.3\% for both methods when Gaussian noise level was 0.1.


\begin{table}[<options>]
\caption{
    The difference between direct measurements ($t_0 \rightarrow t_2$) and two-step measurement ($t_0 \rightarrow t_1$ and $t_1 \rightarrow t_2$). The numbers are reported in percentage and as mean atrophy ($\pm$ standard deviation) across all triplets described in Sec.~\ref{sec:triplet_data}. The mean absolute error ($\pm$ standard deviation) of BVC is also reported.}\label{tab:triplet}
\begin{tabular*}{\tblwidth}{@{}LLL@{}}
\toprule
 Method & Mean ($\pm$ std) $\%$ & Mean Absolute ($\pm$ std) $\%$ \\ 
\midrule
 SIENA &  0.031 ($\pm$ 0.154) & 0.123 ($\pm$ 0.097) \\
 \Method & 0.028 ($\pm$ 0.145) & 0.120 ($\pm$ 0.087) \\
\bottomrule
\end{tabular*}
\end{table}

\begin{figure*}[<options>]
	\centering
            \includegraphics[scale=0.04]{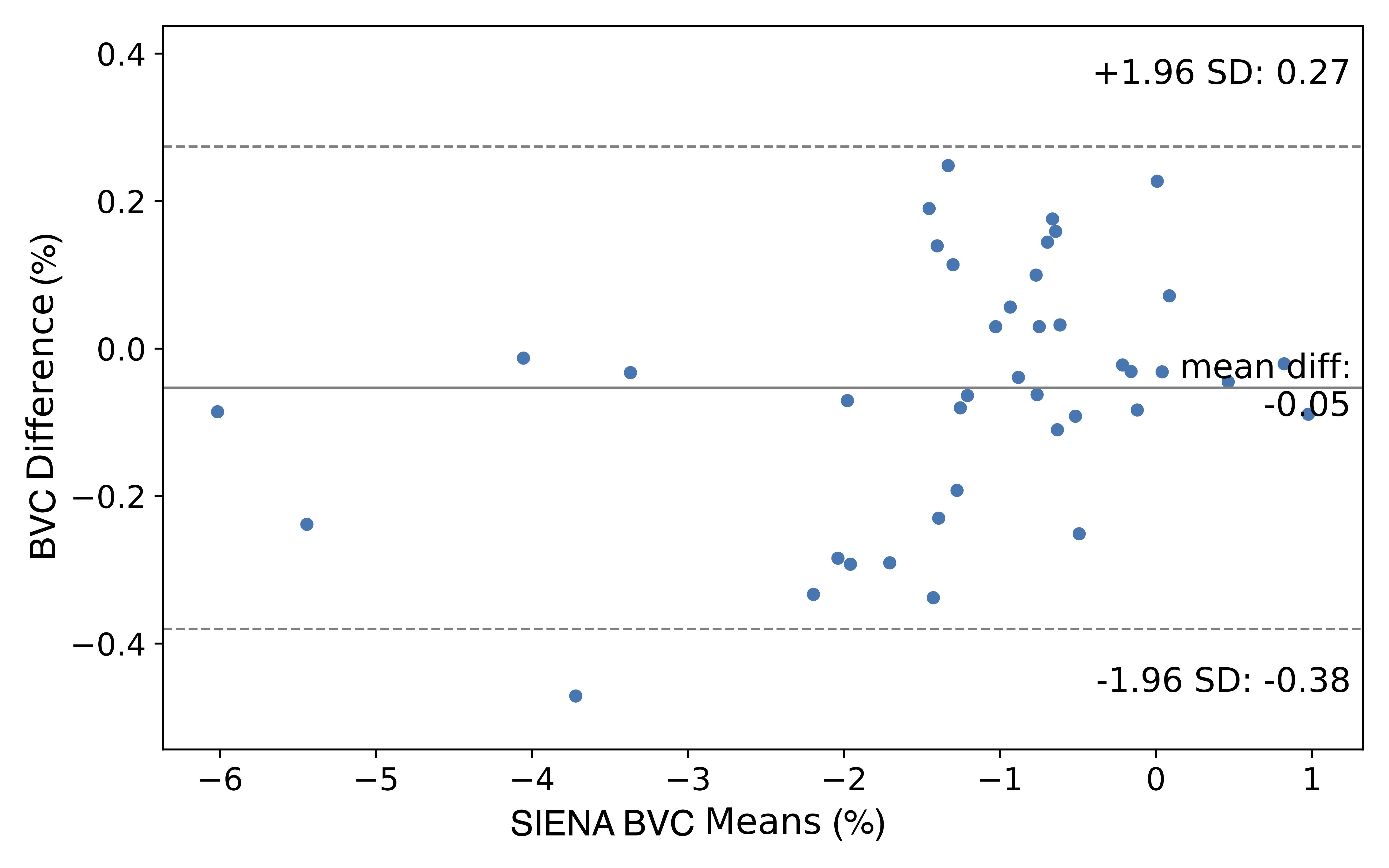}
            \includegraphics[scale=0.04]{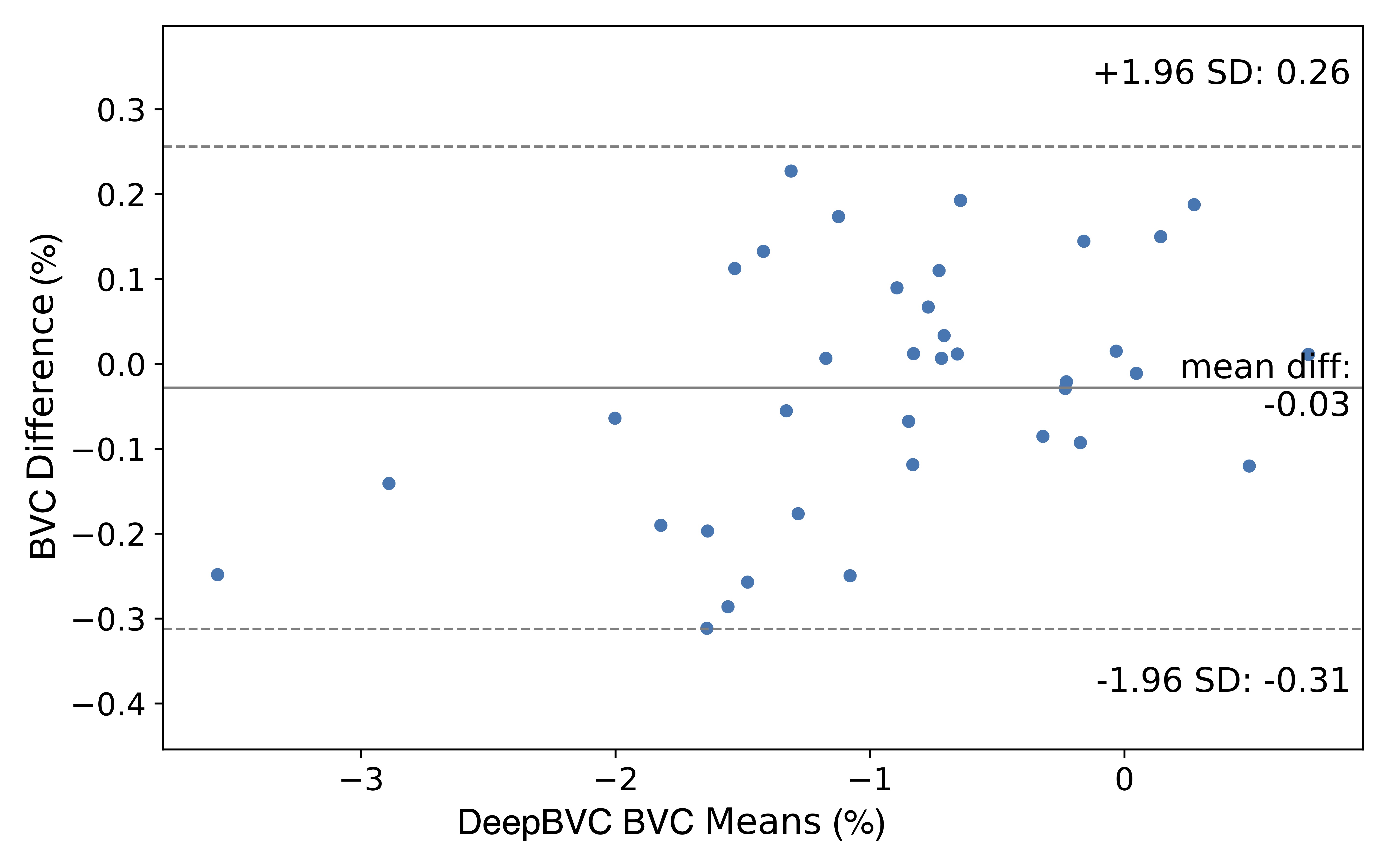}
            \includegraphics[scale=0.04]{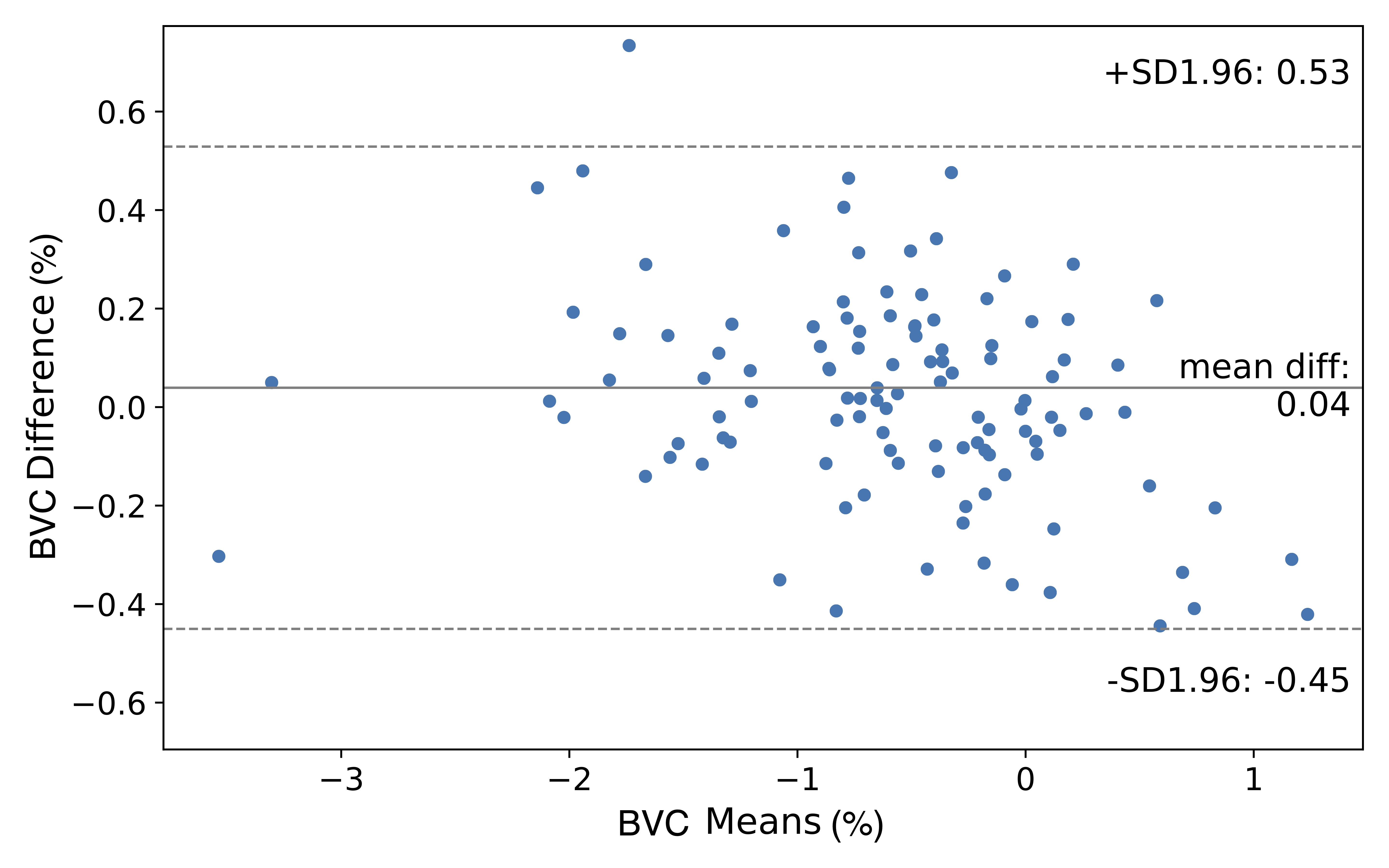}
	  \caption{
   Bland-Altman plots. Difference of direct and indirect measurements for SIENA (top left) and \Method (top right) respectively. The bottom plot compares the measurement difference between SIENA and \Method (Sec.~\ref{sec:triplet_data}).
   }\label{fig:direct-indirect-bland-altman}
\end{figure*}

\begin{figure}
	\centering
            \includegraphics[scale=0.55]{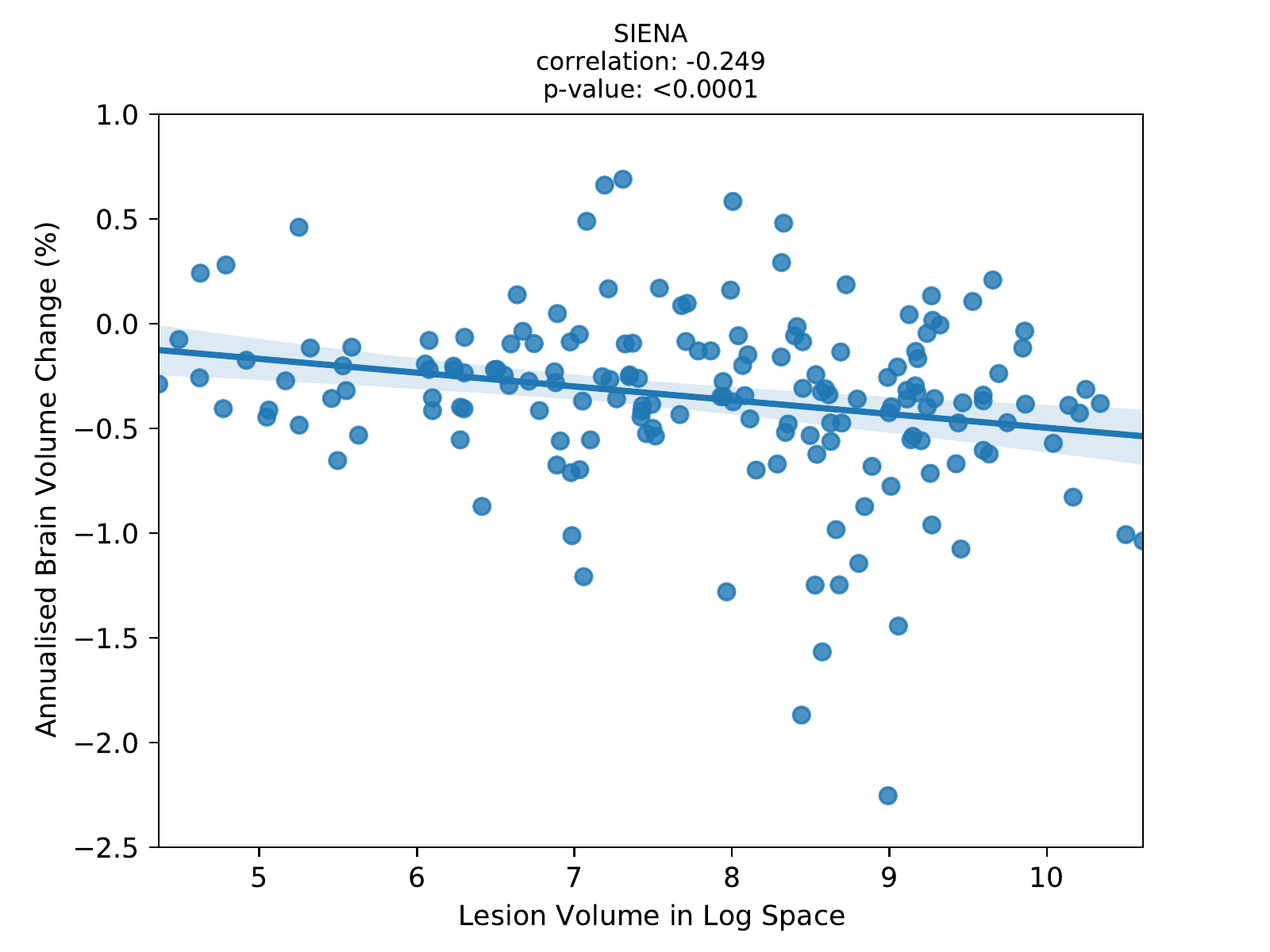}
            \includegraphics[scale=0.55]{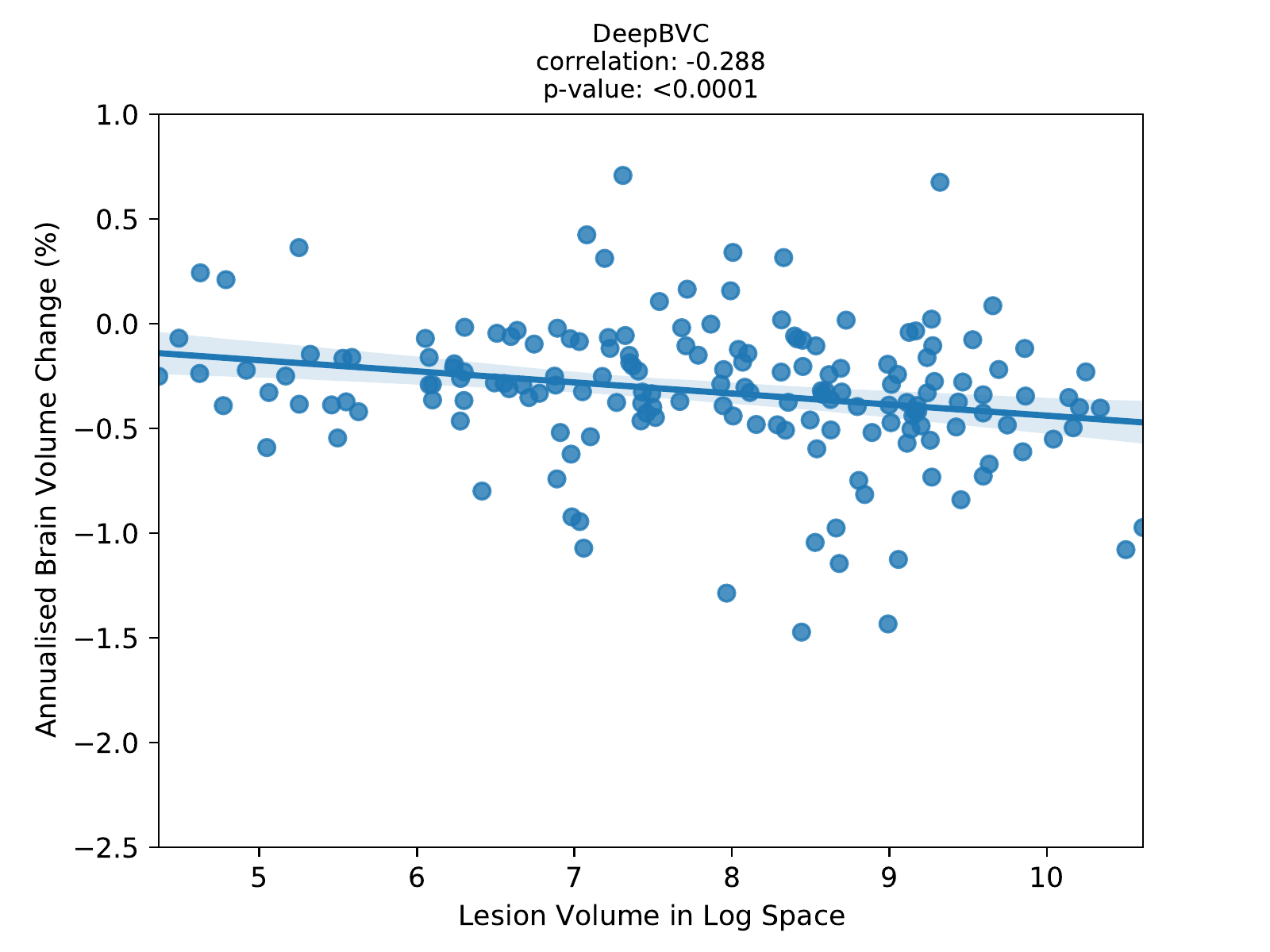}
	  \caption{
    The correlation between annualised brain atrophy rate (over 3-4 years) and baseline lesion volume.
   }\label{fig:correlation-lesion-volume}
\end{figure}

\subsection{Multi-step Consistency with Three Time Points}
Among the 233 subjects involved in this evaluation experiment, the $mean$ ($\pm std$) time interval for $t_0 \rightarrow t_1$ was $1.6 (\pm 0.9)$ years (range $1.0 - 6.9$ years); 
for $t_1 \rightarrow t_2$ , $1.6 (\pm 0.8)$ years (range $1.0 - 6.7$ years); 
and for $t_0 \rightarrow t_2$, $3.2 (\pm 1.2)$ years (range $2.0 - 9.4$ years).

We report the difference in single-step and two-step BVC measurements for the two methods in Tab.~\ref{tab:triplet} and Fig.~\ref{fig:direct-indirect-bland-altman}.

The mean (± standard deviation) for SIENA and \Method was $0.031\% (\pm 0.154\%)$ and $0.028\% (\pm 0.145\%)$ respectively. The mean absolute error (±standard deviation) was $0.123\% (\pm 0.097\%)$ and $0.120\% (\pm 0.087\%)$ respectively. The difference between direct ($t_0 \rightarrow t_2$) and two-step measurements ($t_0$ $\rightarrow$ $t_1$ and $t_1$ $\rightarrow$ $t_2$)  was relatively smaller for our method ($p=0.78$).


When comparing the direct and indirect measurement for each method using Bland-Altman plots, the points are scattered randomly above and below 0 for both SIENA and \Method (Fig.~\ref{fig:direct-indirect-bland-altman}).
The $1.96 SD$ and $-1.96 SD$ for SIENA is $0.27$ and $-0.38$ respectively (approximating the values reported by \cite{smith2002accurate}), whereas the $1.96 SD$ and $-1.96 SD$ for \Method is $0.26$ and $-0.31$ respectively.
\Method was less likely than SIENA to generate an output that suggested brain volume growth (biologically less likely) over time; and there was no obvious bias between two methods. 
For points with large atrophy (brain loss $> 2\%$), most points are between $SD 1.96=0.53$ and $-SD 1.96=-0.45$ with only one point outside this range.



\subsection{Brain T2 Lesion Volume and Brain Volume Change}
We report the correlation between the annualised brain atrophy rate and the total lesion volume at baseline. As shown in Fig.~\ref{fig:correlation-lesion-volume}, our method had a slightly stronger linear correlation with baseline lesion volume ($r_{s}=- 0.288$, $p<0.05$) than PBVC-SIENA ($r_{s}=-0.249$, $p<0.05$). A similar trend was observed on partial correlation controlled for age and disease duration (Tab.~\ref{fig:correlation_lesion_controlled}), with $r_{s}=-0.373 (p<0.05)$ for the deep learning model, and $r_{s}=-0.339 (p<0.05)$ for SIENA.


\subsection{Disability and Brain Volume Change}
\begin{figure*}
	\centering
            \includegraphics[scale=0.06]{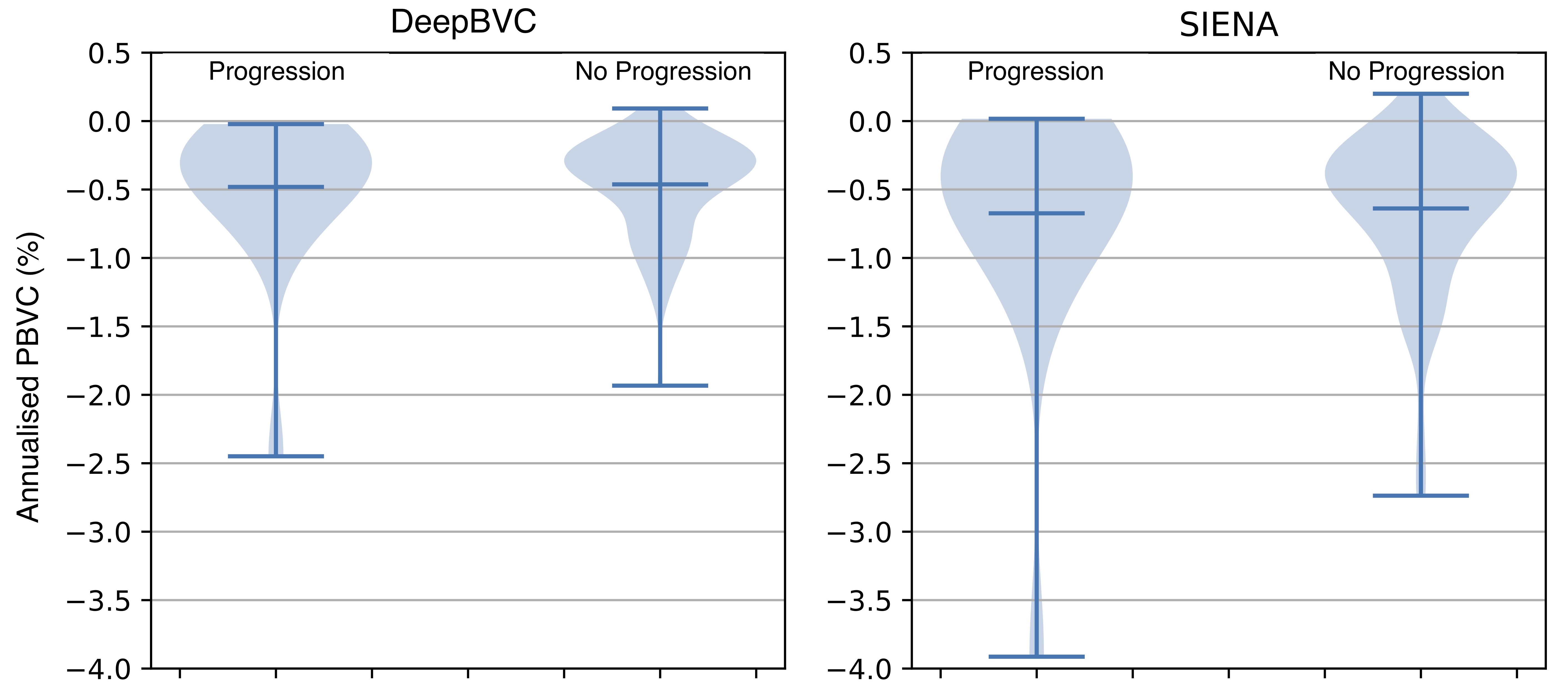}
	  \caption{
    The violin plots of annualised BVC for two groups: sustained EDSS progression and no sustained EDSS progression.
   }\label{fig:correlation-edss}
\end{figure*}

\begin{figure*}
	\centering
            \includegraphics[scale=0.06]{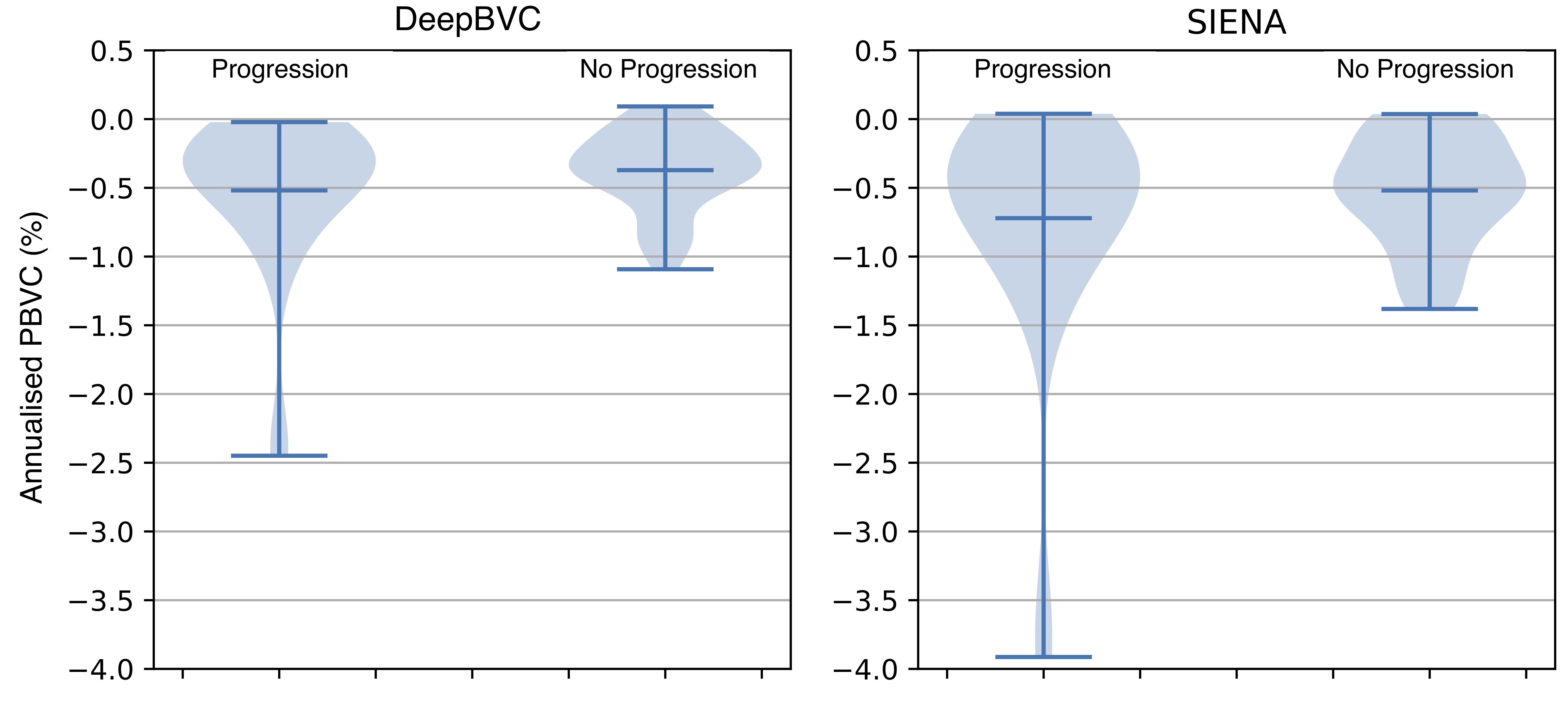}
	  \caption{
    The violin plots of annualised BVC for matched subjects from two groups: sustained EDSS progression and no sustained EDSS progression. The matching is one-to-multiple propensity score matching on age and disease duration.
   }\label{fig:correlation-edss-prop}
\end{figure*}

The annualised BVC distribution is shown in Fig.~\ref{fig:correlation-edss} for subjects with and without sustained EDSS progression. For both \Method and SIENA, the average annualised BVC is slightly larger for the group with sustained EDSS progression (for \Method $p=0.86$, for SIENA $p=0.84$).  The annualised BVC for matched subjects with and without sustained EDSS progression is shown in Fig.~\ref{fig:correlation-edss-prop}. For both \Method and SIENA, the average annualised BVC for the subjects with progression was larger (for \Method $p=0.31$, for SIENA $p=0.35$). For both experiments, there was no significant correlation of disability progression with BVC as estimated by either method.

\begin{table}[<options>]
\caption{
    Correlation (controlled for age and disease duration) between annualised brain atrophy rate (over 3-4 years) and baseline lesion volume.}\label{fig:correlation_lesion_controlled}
\begin{tabular*}{\tblwidth}{LLL}
\toprule
 Method & Correlation $r_{s}$ & p-value \\ 
\midrule
 SIENA &  -0.339 & 0.005 \\
 Ours & -0.373 & 0.002 \\
\bottomrule
\end{tabular*}
\end{table}

\section{Discussion}
Deep learning with pseudo labels has been previously used in the field of neuroimaging. For example, FastSurfer’s deep learning algorithms were trained on outputs generated by the conventional neuroimaging pipelines that underpin Freesurfer~\citep{henschel2020fastsurfer}. The reliability, sensitivity, and time efficiency of FastSurfer is proven to be superior to FreeSurfer~\citep{fischl2012freesurfer}. While FastSurfer and DeepBVC share the concept of using pseudo-labels for training, there are two significant differences. First, the output of FastSurfer is a segmentation mask and is generated by classification, while the output of our method is produced by regression. Second, FastSurfer focuses on designing a novel deep learning network to improve efficiency, whereas our method uses a simple but effective network and regularisation technique to reduce the impact of noise in the pseudo-label.

The data augmentation and consistency regularisation of our method are only used during training; therefore the efficiency of the model is not affected by those techniques. Data augmentation improves the model’s generalisability to unseen data~\citep{zhou2022domain}. In our case, augmentation simulates inconsistent protocols during the acquisition of scan pairs. Therefore, the \Method model is adapted to those types of protocol inconsistency and potentially generalises well to similar data. Though we only included random contrast and spacing anisotropy in training, the model demonstrated improved reliability on test-retest data with noise-related inconsistencies. Consistency regularisation renders the predictions invariant to noise applied to the input, and is widely used in semi-supervised learning and learning from noisy labels~\citep{miyato2018virtual, sajjadi2016regularization, clark2018semi}. Specifically, the incorporation of regularisation into our model maintains an identical brain volume change between scan pairs acquired with either a consistent protocol or a (synthetically generated) inconsistent protocol. These two techniques enable the model to learn clean predictions from noisy pseudo-labels; and improve the estimation reliability, especially for longitudinal scan pairs with an inconsistent acquisition protocol.

In general, smaller BVC on the test-retest data indicates better reliability. As in Tab.~\ref{tab:original_test_retest}, for each of the three subjects, our method estimated a smaller test-retest brain volume difference (range $0.066-0.126\%$) and smaller standard deviations than SIENA (range $0.118-0.351\%$) across sessions ($p=0.03, p=0.08, p<0.001$ for subject 1, 2, 3 respectively). The reliability of \Method also appears to be less sensitive to subject-related factors, based on the results of subjects 1 and 3 from the test-retest dataset. For example, the mean BVC estimated by SIENA for subject 3 was greater than 3 times that of subject 1, whereas estimation by our method was less than 2 times. Based on these observations, the reliability of \Method is superior to SIENA for the test-retest data with a consistent protocol.  We simulated a limited array of protocol inconsistencies commonly observed in clinical practice in the test-retest dataset.  Specifically, we investigated the impact of contrast variance, which may be introduced by changes in head coil or sequence parameters such as TE~\citep{constable1992factors}; bias field, which may relate to spatial variance in coil sensitivity and the interaction between the scanner and the subject~\citep{kim2011bias}; and image resolution, which is determined by scanner/sequence settings. \Method demonstrated superior or at least equivalent performance when compared to SIENA in all scenarios other than inconsistency in the context of bias field. Among the four types of inconsistency tested, contrast and spacing anisotropy variance had the greatest impact on SIENA measurements, followed by Gaussian noise and bias field. \Method showed significantly better performance in the context of contrast and spacing anisotropy pseudo-inconsistencies, despite the lack of spacing anisotropy variation and far less extreme contrast variation in training data augmentation, indicating generalisabilty of the method to unseen scenarios. Surprisingly, Gaussian noise and random bias field did not significantly impact the measurement for SIENA. \Method demonstrates better reliability against most levels of Gaussian noise. However, potentially reflecting the fact that SIENA relies on edge-enhanced image profiles, which are robust to local intensity differences between images.  For \Method, input images are pre-processed with voxel intensity normalisation and bias field alters the distribution of the input voxel intensities after the normalisation step. Although SIENA estimates are more robust to random bias field fluctuations, the largest median BVC for \Method was only $0.29\%$ , a fraction of the magnitude of error introduced into SIENA estimates by other types of inconsistency. In general, \Method is therefore more reliable and less sensitive to protocol inconsistency.

Using patient data from three time points, a smaller difference between one-step and multi-step measurements indicates better consistency. The application of both SIENA and \Method in this experimental paradigm yielded a small difference ($\approx 0.03\%$) between one-step ($t_0 \rightarrow t_2$) and two-step ($t_0 \rightarrow t_1$ and $t_1 \rightarrow t_2$) measurements, indicating high consistency for both methods, marginally in favour of \Method. The difference between one-step and multi-step measurements can also reveal systematic errors; a smaller difference therefore indicates that less effort is required for calibration for studies that involve multiple ($\ge 3$) data timepoints. For both SIENA and \Method, the differences between the two measurements randomly scattered above and below 0 in the Bland Altman comparison (Fig.\ref{fig:direct-indirect-bland-altman}), suggesting that there were no significant accumulative or systematic measurement errors. For both SIENA and \Method, a high agreement between the direct and indirect measurements was observed ($+1.96 SD=0.27$ and $-1.96 SD=-0.38$ for SIENA; $+1.96 SD=0.26$ and $-1.96 SD=-0.31$ for \Method), especially for the subjects with large mean atrophy.

Correlation experiments (Fig.~\ref{fig:correlation-lesion-volume}) illustrated that \Method estimates of BVC had a marginally stronger linear correlation with baseline brain lesion volume (\Method: $r=0.288$; SIENA: $r=0.249$), an advantage that was retained when confounding variables (age and disease duration) were controlled (\Method: $r=0.373$; SIENA: $r=0.339$). These findings suggest fewer subjects may be required to power group-level studies that use our tool to estimate BVC as an endpoint.

EDSS experiments demonstrate that a higher annualised BVC was observed amongst EDSS progressors compared with non-progressors group, but the difference was not significant for either \Method ($p=0.86$) or SIENA ($p=0.84$). Similarly, analysis of propensity score matched subjects showed higher, but not significant, annualised BVC amongst EDSS progressors for both methods (DeepBVC: $p=0.31$; SIENA: $p=0.35$) The magnitude of group level differences in annualised BVC between EDSS progressors and non-progressors wereless than in previous reports~\cite{rudick2000brain, bermel2006measurement}. Patient populations in these earlier studies differed from the modern MS cohort, in which the majority of patients are treated with high efficacy disease modifying agents (that are known to reduce BVC), studied in the present work.  Additionally, these studies employed different measures to determine brain volume (change), such as brain parenchymal fraction and normalised whole brain grey-matter volume. 

\textbf{Study limitations and future directions.}
While our experiments suggest that DeepBVC more consistently and reliably estimates BVC than the classical tool, SIENA, in several scenarios, there are a number of limitations.

First, model training requires pseudo-labels from SIENA. While the use of pseudo- labels generates improved performance, the overall framework and concept follow the principles of the classical method, namely the requirements for co-registration of baseline and follow-up scans, segmentation of the brain to find edge points, and a calibration step for the final volume change estimate. As a consequence, our experimental results may be confounded by errors propagated from each pre-processing step. For example, the registration step potentially changes the scale and skew of the brain image, which can in turn affect the final BVC estimation. Additionally, lesion inpainting changes the image context and affects brain edge point segmentation, impacting the edge locations at which voxel-wise atrophy/growth is subsequently estimated. Future pipeline optimisation, in which each step is integrated as a component of the learning process, may mitigate this cascading effect and enhance the performance of deep learning based solutions for BVC. 

Second, we simulated four protocol inconsistencies and independently tested their impact on BVC estimates using \Method. In real-world clinical imaging acquisitions, protocol inconsistencies are more complex, overlapping and generated by different and, at times, multiple sources. Decomposing these inconsistencies into isolated categories is challenging.  Similarly, the availability of inconsistently acquired scans, acquired back-to-back on the same subject, would be required to complete the test-retest study in the absence of simulated data.

Third, it is challenging to validate the actual measurement accuracy of tools such as SIENA or \Method, because the ground truth BVC is unknown, other than for test-retest subjects in whom brain volume should essentially remain static when scans are acquired back to back, thereby avoiding changes in hydration or diurnal factors that could impact brain volume. Although \Method did not inappropriately detect BVC in the test-retest cohort, this does not necessarily demonstrate capacity to accurately determine atrophy estimates in cases with true brain tissue volume differences. In this regard, BVC estimates from SIENA are noisy: they can be used to produce pseudo-labels for the training phase, but they cannot be used as ground truth during the validation phase, particularly without rigorous manual quality control checks. We therefore resorted to validating \Method using proxy techniques (Sec.~\ref{sec:experimenta_design}) in the absence of ‘clean’ labels for testing model performance.  While these methods were comprehensive and approximated real world imaging scenarios, multiple time-consuming steps in the experimental pipeline hindered rapid model development.

\section{Conclusions}
We demonstrate that a deep learning model trained for whole BVC estimation with pseudo-labels derived from SIENA can achieve better performance in terms of consistency, invariance to protocol change, and correlation between BVC and baseline lesion volume in a cohort of subjects with MS. Brain atrophy is a common endpoint in MS clinical trials that will become more relevant as neuroprotective and pro-reparative therapies are developed. Similarly, there is a need for robust monitoring of brain atrophy in neurodegenerative disease, the imperative for which has been heightened by the recent advent of disease modifying therapies in this patient population.  DeepBVC is a fast and robust method for estimating brain atrophy that may have particular application in both clinical trials and precision medicine.

\section*{Declaration of Interests}
Geng Zhan is a part-time employee at the Sydney Neuroimaging Analysis Centre. Dongang Wang is a part-time employee at the Sydney Neuroimaging Analysis Centre. Mariano Cabezas has nothing to disclose. Lei Bai has nothing to disclose. Kain Kyle is a part-time employee at the Sydney Neuroimaging Analysis Centre. Wanli Ouyang has nothing to disclose. Michael Barnett has received institutional support for research, speaking and/or participation in advisory boards for Biogen, Merck, Novartis, Roche and Sanofi Genzyme, and is a research consultant to RxPx and research director for the Sydney Neuroimaging Analysis Centre. Chenyu Wang is a part-time employee at the Sydney Neuroimaging Analysis Centre.








\printcredits

\bibliographystyle{cas-model2-names}

\bibliography{cas-dc-template}

\begin{thebibliography}{66}
\expandafter\ifx\csname natexlab\endcsname\relax\def\natexlab#1{#1}\fi
\providecommand{\url}[1]{\texttt{#1}}
\providecommand{\href}[2]{#2}
\providecommand{\path}[1]{#1}
\providecommand{\DOIprefix}{doi:}
\providecommand{\ArXivprefix}{arXiv:}
\providecommand{\URLprefix}{URL: }
\providecommand{\Pubmedprefix}{pmid:}
\providecommand{\doi}[1]{\href{http://dx.doi.org/#1}{\path{#1}}}
\providecommand{\Pubmed}[1]{\href{pmid:#1}{\path{#1}}}
\providecommand{\bibinfo}[2]{#2}
\ifx\xfnm\relax \def\xfnm[#1]{\unskip,\space#1}\fi
\bibitem[{Beer et~al.(2020)Beer, Tustison, Cook, Davatzikos, Sheline,
  Shinohara, Linn, Initiative et~al.}]{beer2020longitudinal}
\bibinfo{author}{Beer, J.C.}, \bibinfo{author}{Tustison, N.J.},
  \bibinfo{author}{Cook, P.A.}, \bibinfo{author}{Davatzikos, C.},
  \bibinfo{author}{Sheline, Y.I.}, \bibinfo{author}{Shinohara, R.T.},
  \bibinfo{author}{Linn, K.A.}, \bibinfo{author}{Initiative, A.D.N.}, et~al.,
  \bibinfo{year}{2020}.
\newblock \bibinfo{title}{Longitudinal combat: A method for harmonizing
  longitudinal multi-scanner imaging data}.
\newblock \bibinfo{journal}{Neuroimage} \bibinfo{volume}{220},
  \bibinfo{pages}{117129}.
\bibitem[{Bermel and Bakshi(2006)}]{bermel2006measurement}
\bibinfo{author}{Bermel, R.A.}, \bibinfo{author}{Bakshi, R.},
  \bibinfo{year}{2006}.
\newblock \bibinfo{title}{The measurement and clinical relevance of brain
  atrophy in multiple sclerosis}.
\newblock \bibinfo{journal}{The Lancet Neurology} \bibinfo{volume}{5},
  \bibinfo{pages}{158--170}.
\bibitem[{Bermel et~al.(2003)Bermel, Sharma, Tjoa, Puli and
  Bakshi}]{bermel2003semiautomated}
\bibinfo{author}{Bermel, R.A.}, \bibinfo{author}{Sharma, J.},
  \bibinfo{author}{Tjoa, C.W.}, \bibinfo{author}{Puli, S.R.},
  \bibinfo{author}{Bakshi, R.}, \bibinfo{year}{2003}.
\newblock \bibinfo{title}{A semiautomated measure of whole-brain atrophy in
  multiple sclerosis}.
\newblock \bibinfo{journal}{Journal of the neurological sciences}
  \bibinfo{volume}{208}, \bibinfo{pages}{57--65}.
\bibitem[{Cadavid et~al.(2017)Cadavid, Balcer, Galetta, Aktas, Ziemssen,
  Vanopdenbosch, Frederiksen, Skeen, Jaffe, Butzkueven
  et~al.}]{cadavid2017safety}
\bibinfo{author}{Cadavid, D.}, \bibinfo{author}{Balcer, L.},
  \bibinfo{author}{Galetta, S.}, \bibinfo{author}{Aktas, O.},
  \bibinfo{author}{Ziemssen, T.}, \bibinfo{author}{Vanopdenbosch, L.},
  \bibinfo{author}{Frederiksen, J.}, \bibinfo{author}{Skeen, M.},
  \bibinfo{author}{Jaffe, G.J.}, \bibinfo{author}{Butzkueven, H.}, et~al.,
  \bibinfo{year}{2017}.
\newblock \bibinfo{title}{Safety and efficacy of opicinumab in acute optic
  neuritis (renew): a randomised, placebo-controlled, phase 2 trial}.
\newblock \bibinfo{journal}{The Lancet Neurology} \bibinfo{volume}{16},
  \bibinfo{pages}{189--199}.
\bibitem[{Chen et~al.(2004)Chen, Reiman, Alexander, Bandy, Renaut, Crum, Fox
  and Rossor}]{chen2004automated}
\bibinfo{author}{Chen, K.}, \bibinfo{author}{Reiman, E.M.},
  \bibinfo{author}{Alexander, G.E.}, \bibinfo{author}{Bandy, D.},
  \bibinfo{author}{Renaut, R.}, \bibinfo{author}{Crum, W.R.},
  \bibinfo{author}{Fox, N.C.}, \bibinfo{author}{Rossor, M.N.},
  \bibinfo{year}{2004}.
\newblock \bibinfo{title}{An automated algorithm for the computation of brain
  volume change from sequential mris using an iterative principal component
  analysis and its evaluation for the assessment of whole-brain atrophy rates
  in patients with probable alzheimer's disease}.
\newblock \bibinfo{journal}{Neuroimage} \bibinfo{volume}{22},
  \bibinfo{pages}{134--143}.
\bibitem[{{\c{C}}i{\c{c}}ek et~al.(2016){\c{C}}i{\c{c}}ek, Abdulkadir,
  Lienkamp, Brox and Ronneberger}]{cciccek20163d}
\bibinfo{author}{{\c{C}}i{\c{c}}ek, {\"O}.}, \bibinfo{author}{Abdulkadir, A.},
  \bibinfo{author}{Lienkamp, S.S.}, \bibinfo{author}{Brox, T.},
  \bibinfo{author}{Ronneberger, O.}, \bibinfo{year}{2016}.
\newblock \bibinfo{title}{3d u-net: learning dense volumetric segmentation from
  sparse annotation}, in: \bibinfo{booktitle}{International conference on
  medical image computing and computer-assisted intervention},
  \bibinfo{organization}{Springer}. pp. \bibinfo{pages}{424--432}.
\bibitem[{Clark et~al.(2018)Clark, Luong, Manning and Le}]{clark2018semi}
\bibinfo{author}{Clark, K.}, \bibinfo{author}{Luong, M.T.},
  \bibinfo{author}{Manning, C.D.}, \bibinfo{author}{Le, Q.V.},
  \bibinfo{year}{2018}.
\newblock \bibinfo{title}{Semi-supervised sequence modeling with cross-view
  training}.
\newblock \bibinfo{journal}{arXiv preprint arXiv:1809.08370} .
\bibitem[{Collins et~al.(2001)Collins, Montagnat, Zijdenbos, Evans and
  Arnold}]{collins2001automated}
\bibinfo{author}{Collins, D.L.}, \bibinfo{author}{Montagnat, J.},
  \bibinfo{author}{Zijdenbos, A.P.}, \bibinfo{author}{Evans, A.C.},
  \bibinfo{author}{Arnold, D.L.}, \bibinfo{year}{2001}.
\newblock \bibinfo{title}{Automated estimation of brain volume in multiple
  sclerosis with biccr}, in: \bibinfo{booktitle}{Biennial International
  Conference on Information Processing in Medical Imaging},
  \bibinfo{organization}{Springer}. pp. \bibinfo{pages}{141--147}.
\bibitem[{Constable et~al.(1992)Constable, Anderson, Zhong and
  Gore}]{constable1992factors}
\bibinfo{author}{Constable, R.}, \bibinfo{author}{Anderson, A.},
  \bibinfo{author}{Zhong, J.}, \bibinfo{author}{Gore, J.},
  \bibinfo{year}{1992}.
\newblock \bibinfo{title}{Factors influencing contrast in fast spin-echo mr
  imaging}.
\newblock \bibinfo{journal}{Magnetic resonance imaging} \bibinfo{volume}{10},
  \bibinfo{pages}{497--511}.
\bibitem[{De~Stefano et~al.(2014)De~Stefano, Airas, Grigoriadis, Mattle,
  O’Riordan, Oreja-Guevara, Sellebjerg, Stankoff, Walczak, Wiendl
  et~al.}]{de2014clinical}
\bibinfo{author}{De~Stefano, N.}, \bibinfo{author}{Airas, L.},
  \bibinfo{author}{Grigoriadis, N.}, \bibinfo{author}{Mattle, H.P.},
  \bibinfo{author}{O’Riordan, J.}, \bibinfo{author}{Oreja-Guevara, C.},
  \bibinfo{author}{Sellebjerg, F.}, \bibinfo{author}{Stankoff, B.},
  \bibinfo{author}{Walczak, A.}, \bibinfo{author}{Wiendl, H.}, et~al.,
  \bibinfo{year}{2014}.
\newblock \bibinfo{title}{Clinical relevance of brain volume measures in
  multiple sclerosis}.
\newblock \bibinfo{journal}{CNS drugs} \bibinfo{volume}{28},
  \bibinfo{pages}{147--156}.
\bibitem[{De~Stefano et~al.(2016)De~Stefano, Stromillo, Giorgio, Bartolozzi,
  Battaglini, Baldini, Portaccio, Amato and Sormani}]{de2016establishing}
\bibinfo{author}{De~Stefano, N.}, \bibinfo{author}{Stromillo, M.L.},
  \bibinfo{author}{Giorgio, A.}, \bibinfo{author}{Bartolozzi, M.L.},
  \bibinfo{author}{Battaglini, M.}, \bibinfo{author}{Baldini, M.},
  \bibinfo{author}{Portaccio, E.}, \bibinfo{author}{Amato, M.P.},
  \bibinfo{author}{Sormani, M.P.}, \bibinfo{year}{2016}.
\newblock \bibinfo{title}{Establishing pathological cut-offs of brain atrophy
  rates in multiple sclerosis}.
\newblock \bibinfo{journal}{Journal of Neurology, Neurosurgery \& Psychiatry}
  \bibinfo{volume}{87}, \bibinfo{pages}{93--99}.
\bibitem[{Dewey et~al.(2019)Dewey, Zhao, Reinhold, Carass, Fitzgerald,
  Sotirchos, Saidha, Oh, Pham, Calabresi et~al.}]{dewey2019deepharmony}
\bibinfo{author}{Dewey, B.E.}, \bibinfo{author}{Zhao, C.},
  \bibinfo{author}{Reinhold, J.C.}, \bibinfo{author}{Carass, A.},
  \bibinfo{author}{Fitzgerald, K.C.}, \bibinfo{author}{Sotirchos, E.S.},
  \bibinfo{author}{Saidha, S.}, \bibinfo{author}{Oh, J.},
  \bibinfo{author}{Pham, D.L.}, \bibinfo{author}{Calabresi, P.A.}, et~al.,
  \bibinfo{year}{2019}.
\newblock \bibinfo{title}{Deepharmony: A deep learning approach to contrast
  harmonization across scanner changes}.
\newblock \bibinfo{journal}{Magnetic resonance imaging} \bibinfo{volume}{64},
  \bibinfo{pages}{160--170}.
\bibitem[{Dewey et~al.(2020)Dewey, Zuo, Carass, He, Liu, Mowry, Newsome, Oh,
  Calabresi and Prince}]{dewey2020disentangled}
\bibinfo{author}{Dewey, B.E.}, \bibinfo{author}{Zuo, L.},
  \bibinfo{author}{Carass, A.}, \bibinfo{author}{He, Y.}, \bibinfo{author}{Liu,
  Y.}, \bibinfo{author}{Mowry, E.M.}, \bibinfo{author}{Newsome, S.},
  \bibinfo{author}{Oh, J.}, \bibinfo{author}{Calabresi, P.A.},
  \bibinfo{author}{Prince, J.L.}, \bibinfo{year}{2020}.
\newblock \bibinfo{title}{A disentangled latent space for cross-site mri
  harmonization}, in: \bibinfo{booktitle}{International Conference on Medical
  Image Computing and Computer-Assisted Intervention},
  \bibinfo{organization}{Springer}. pp. \bibinfo{pages}{720--729}.
\bibitem[{Duffy et~al.(2018)Duffy, Zhang, Tang, Zhao, Law, Toga and
  Kim}]{duffy2018retrospective}
\bibinfo{author}{Duffy, B.A.}, \bibinfo{author}{Zhang, W.},
  \bibinfo{author}{Tang, H.}, \bibinfo{author}{Zhao, L.}, \bibinfo{author}{Law,
  M.}, \bibinfo{author}{Toga, A.W.}, \bibinfo{author}{Kim, H.},
  \bibinfo{year}{2018}.
\newblock \bibinfo{title}{Retrospective correction of motion artifact affected
  structural mri images using deep learning of simulated motion} .
\bibitem[{Field(2013)}]{field2013discovering}
\bibinfo{author}{Field, A.}, \bibinfo{year}{2013}.
\newblock \bibinfo{title}{Discovering statistics using IBM SPSS statistics}.
\newblock \bibinfo{publisher}{sage}.
\bibitem[{Filippi et~al.(2004)Filippi, Rovaris, Inglese, Barkhof, De~Stefano,
  Smith, Comi, Group et~al.}]{filippi2004interferon}
\bibinfo{author}{Filippi, M.}, \bibinfo{author}{Rovaris, M.},
  \bibinfo{author}{Inglese, M.}, \bibinfo{author}{Barkhof, F.},
  \bibinfo{author}{De~Stefano, N.}, \bibinfo{author}{Smith, S.},
  \bibinfo{author}{Comi, G.}, \bibinfo{author}{Group, E.S.}, et~al.,
  \bibinfo{year}{2004}.
\newblock \bibinfo{title}{Interferon beta-1a for brain tissue loss in patients
  at presentation with syndromes suggestive of multiple sclerosis: a
  randomised, double-blind, placebo-controlled trial}.
\newblock \bibinfo{journal}{The Lancet} \bibinfo{volume}{364},
  \bibinfo{pages}{1489--1496}.
\bibitem[{Fischl(2012)}]{fischl2012freesurfer}
\bibinfo{author}{Fischl, B.}, \bibinfo{year}{2012}.
\newblock \bibinfo{title}{Freesurfer}.
\newblock \bibinfo{journal}{Neuroimage} \bibinfo{volume}{62},
  \bibinfo{pages}{774--781}.
\bibitem[{Freeborough and Fox(1997)}]{freeborough1997boundary}
\bibinfo{author}{Freeborough, P.A.}, \bibinfo{author}{Fox, N.C.},
  \bibinfo{year}{1997}.
\newblock \bibinfo{title}{The boundary shift integral: an accurate and robust
  measure of cerebral volume changes from registered repeat mri}.
\newblock \bibinfo{journal}{IEEE transactions on medical imaging}
  \bibinfo{volume}{16}, \bibinfo{pages}{623--629}.
\bibitem[{Friston(2003)}]{friston2003statistical}
\bibinfo{author}{Friston, K.J.}, \bibinfo{year}{2003}.
\newblock \bibinfo{title}{Statistical parametric mapping}, in:
  \bibinfo{booktitle}{Neuroscience databases}. \bibinfo{publisher}{Springer},
  pp. \bibinfo{pages}{237--250}.
\bibitem[{Garcia-Dias et~al.(2020)Garcia-Dias, Scarpazza, Baecker, Vieira,
  Pinaya, Corvin, Redolfi, Nelson, Crespo-Facorro, McDonald
  et~al.}]{garcia2020neuroharmony}
\bibinfo{author}{Garcia-Dias, R.}, \bibinfo{author}{Scarpazza, C.},
  \bibinfo{author}{Baecker, L.}, \bibinfo{author}{Vieira, S.},
  \bibinfo{author}{Pinaya, W.H.}, \bibinfo{author}{Corvin, A.},
  \bibinfo{author}{Redolfi, A.}, \bibinfo{author}{Nelson, B.},
  \bibinfo{author}{Crespo-Facorro, B.}, \bibinfo{author}{McDonald, C.}, et~al.,
  \bibinfo{year}{2020}.
\newblock \bibinfo{title}{Neuroharmony: A new tool for harmonizing volumetric
  mri data from unseen scanners}.
\newblock \bibinfo{journal}{Neuroimage} \bibinfo{volume}{220}.
\bibitem[{Goodfellow et~al.(2014)Goodfellow, Shlens and
  Szegedy}]{goodfellow2014explaining}
\bibinfo{author}{Goodfellow, I.J.}, \bibinfo{author}{Shlens, J.},
  \bibinfo{author}{Szegedy, C.}, \bibinfo{year}{2014}.
\newblock \bibinfo{title}{Explaining and harnessing adversarial examples}.
\newblock \bibinfo{journal}{arXiv preprint arXiv:1412.6572} .
\bibitem[{Hajnal et~al.(1995)Hajnal, Saeed, Oatridge, Williams, Young and
  Bydder}]{hajnal1995detection}
\bibinfo{author}{Hajnal, J.V.}, \bibinfo{author}{Saeed, N.},
  \bibinfo{author}{Oatridge, A.}, \bibinfo{author}{Williams, E.J.},
  \bibinfo{author}{Young, I.R.}, \bibinfo{author}{Bydder, G.M.},
  \bibinfo{year}{1995}.
\newblock \bibinfo{title}{Detection of subtle brain changes using subvoxel
  registration and subtraction of serial mr images.}
\newblock \bibinfo{journal}{Journal of computer assisted tomography}
  \bibinfo{volume}{19}, \bibinfo{pages}{677--691}.
\bibitem[{He et~al.(2016)He, Zhang, Ren and Sun}]{he2016deep}
\bibinfo{author}{He, K.}, \bibinfo{author}{Zhang, X.}, \bibinfo{author}{Ren,
  S.}, \bibinfo{author}{Sun, J.}, \bibinfo{year}{2016}.
\newblock \bibinfo{title}{Deep residual learning for image recognition}, in:
  \bibinfo{booktitle}{Proceedings of the IEEE conference on computer vision and
  pattern recognition}, pp. \bibinfo{pages}{770--778}.
\bibitem[{Henschel et~al.(2020)Henschel, Conjeti, Estrada, Diers, Fischl and
  Reuter}]{henschel2020fastsurfer}
\bibinfo{author}{Henschel, L.}, \bibinfo{author}{Conjeti, S.},
  \bibinfo{author}{Estrada, S.}, \bibinfo{author}{Diers, K.},
  \bibinfo{author}{Fischl, B.}, \bibinfo{author}{Reuter, M.},
  \bibinfo{year}{2020}.
\newblock \bibinfo{title}{Fastsurfer-a fast and accurate deep learning based
  neuroimaging pipeline}.
\newblock \bibinfo{journal}{NeuroImage} \bibinfo{volume}{219},
  \bibinfo{pages}{117012}.
\bibitem[{Higaki et~al.(2019)Higaki, Nakamura, Tatsugami, Nakaura and
  Awai}]{higaki2019improvement}
\bibinfo{author}{Higaki, T.}, \bibinfo{author}{Nakamura, Y.},
  \bibinfo{author}{Tatsugami, F.}, \bibinfo{author}{Nakaura, T.},
  \bibinfo{author}{Awai, K.}, \bibinfo{year}{2019}.
\newblock \bibinfo{title}{Improvement of image quality at ct and mri using deep
  learning}.
\newblock \bibinfo{journal}{Japanese journal of radiology}
  \bibinfo{volume}{37}, \bibinfo{pages}{73--80}.
\bibitem[{Holland et~al.(2011)Holland, Dale, Initiative
  et~al.}]{holland2011nonlinear}
\bibinfo{author}{Holland, D.}, \bibinfo{author}{Dale, A.M.},
  \bibinfo{author}{Initiative, A.D.N.}, et~al., \bibinfo{year}{2011}.
\newblock \bibinfo{title}{Nonlinear registration of longitudinal images and
  measurement of change in regions of interest}.
\newblock \bibinfo{journal}{Medical image analysis} \bibinfo{volume}{15},
  \bibinfo{pages}{489--497}.
\bibitem[{Horsfield et~al.(2003)Horsfield, Rovaris, Rocca, Rossi, Benedict,
  Filippi and Bakshi}]{horsfield2003whole}
\bibinfo{author}{Horsfield, M.}, \bibinfo{author}{Rovaris, M.},
  \bibinfo{author}{Rocca, M.}, \bibinfo{author}{Rossi, P.},
  \bibinfo{author}{Benedict, R.}, \bibinfo{author}{Filippi, M.},
  \bibinfo{author}{Bakshi, R.}, \bibinfo{year}{2003}.
\newblock \bibinfo{title}{Whole-brain atrophy in multiple sclerosis measured by
  two segmentation processes from various mri sequences}.
\newblock \bibinfo{journal}{Journal of the neurological sciences}
  \bibinfo{volume}{216}, \bibinfo{pages}{169--177}.
\bibitem[{Jack~Jr et~al.(2008)Jack~Jr, Bernstein, Fox, Thompson, Alexander,
  Harvey, Borowski, Britson, L.~Whitwell, Ward et~al.}]{jack2008alzheimer}
\bibinfo{author}{Jack~Jr, C.R.}, \bibinfo{author}{Bernstein, M.A.},
  \bibinfo{author}{Fox, N.C.}, \bibinfo{author}{Thompson, P.},
  \bibinfo{author}{Alexander, G.}, \bibinfo{author}{Harvey, D.},
  \bibinfo{author}{Borowski, B.}, \bibinfo{author}{Britson, P.J.},
  \bibinfo{author}{L.~Whitwell, J.}, \bibinfo{author}{Ward, C.}, et~al.,
  \bibinfo{year}{2008}.
\newblock \bibinfo{title}{The alzheimer's disease neuroimaging initiative
  (adni): Mri methods}.
\newblock \bibinfo{journal}{Journal of Magnetic Resonance Imaging: An Official
  Journal of the International Society for Magnetic Resonance in Medicine}
  \bibinfo{volume}{27}, \bibinfo{pages}{685--691}.
\bibitem[{Jacobsen et~al.(2014)Jacobsen, Hagemeier, Myhr, Nyland, Lode,
  Bergsland, Ramasamy, Dalaker, Larsen, Farbu et~al.}]{jacobsen2014brain}
\bibinfo{author}{Jacobsen, C.}, \bibinfo{author}{Hagemeier, J.},
  \bibinfo{author}{Myhr, K.M.}, \bibinfo{author}{Nyland, H.},
  \bibinfo{author}{Lode, K.}, \bibinfo{author}{Bergsland, N.},
  \bibinfo{author}{Ramasamy, D.P.}, \bibinfo{author}{Dalaker, T.O.},
  \bibinfo{author}{Larsen, J.P.}, \bibinfo{author}{Farbu, E.}, et~al.,
  \bibinfo{year}{2014}.
\newblock \bibinfo{title}{Brain atrophy and disability progression in multiple
  sclerosis patients: a 10-year follow-up study}.
\newblock \bibinfo{journal}{Journal of Neurology, Neurosurgery \& Psychiatry}
  \bibinfo{volume}{85}, \bibinfo{pages}{1109--1115}.
\bibitem[{Jenkinson et~al.(2005)Jenkinson, Pechaud, Smith
  et~al.}]{jenkinson2005bet2}
\bibinfo{author}{Jenkinson, M.}, \bibinfo{author}{Pechaud, M.},
  \bibinfo{author}{Smith, S.}, et~al., \bibinfo{year}{2005}.
\newblock \bibinfo{title}{Bet2: Mr-based estimation of brain, skull and scalp
  surfaces}, in: \bibinfo{booktitle}{Eleventh annual meeting of the
  organization for human brain mapping}, \bibinfo{organization}{Toronto.}. p.
  \bibinfo{pages}{167}.
\bibitem[{Kalincik et~al.(2015)Kalincik, Cutter, Spelman, Jokubaitis, Havrdova,
  Horakova, Trojano, Izquierdo, Girard, Duquette et~al.}]{kalincik2015defining}
\bibinfo{author}{Kalincik, T.}, \bibinfo{author}{Cutter, G.},
  \bibinfo{author}{Spelman, T.}, \bibinfo{author}{Jokubaitis, V.},
  \bibinfo{author}{Havrdova, E.}, \bibinfo{author}{Horakova, D.},
  \bibinfo{author}{Trojano, M.}, \bibinfo{author}{Izquierdo, G.},
  \bibinfo{author}{Girard, M.}, \bibinfo{author}{Duquette, P.}, et~al.,
  \bibinfo{year}{2015}.
\newblock \bibinfo{title}{Defining reliable disability outcomes in multiple
  sclerosis}.
\newblock \bibinfo{journal}{Brain} \bibinfo{volume}{138},
  \bibinfo{pages}{3287--3298}.
\bibitem[{Kim et~al.(2011)Kim, Habas, Rajagopalan, Scott, Corbett-Detig,
  Rousseau, Barkovich, Glenn and Studholme}]{kim2011bias}
\bibinfo{author}{Kim, K.}, \bibinfo{author}{Habas, P.A.},
  \bibinfo{author}{Rajagopalan, V.}, \bibinfo{author}{Scott, J.A.},
  \bibinfo{author}{Corbett-Detig, J.M.}, \bibinfo{author}{Rousseau, F.},
  \bibinfo{author}{Barkovich, A.J.}, \bibinfo{author}{Glenn, O.A.},
  \bibinfo{author}{Studholme, C.}, \bibinfo{year}{2011}.
\newblock \bibinfo{title}{Bias field inconsistency correction of
  motion-scattered multislice mri for improved 3d image reconstruction}.
\newblock \bibinfo{journal}{IEEE transactions on medical imaging}
  \bibinfo{volume}{30}, \bibinfo{pages}{1704--1712}.
\bibitem[{Kingma and Ba(2014)}]{kingma2014adam}
\bibinfo{author}{Kingma, D.P.}, \bibinfo{author}{Ba, J.}, \bibinfo{year}{2014}.
\newblock \bibinfo{title}{Adam: A method for stochastic optimization}.
\newblock \bibinfo{journal}{arXiv preprint arXiv:1412.6980} .
\bibitem[{Learned-miller and Ahammad(2004)}]{learned2004joint}
\bibinfo{author}{Learned-miller, E.}, \bibinfo{author}{Ahammad, P.},
  \bibinfo{year}{2004}.
\newblock \bibinfo{title}{Joint mri bias removal using entropy minimization
  across images}.
\newblock \bibinfo{journal}{Advances in Neural Information Processing Systems}
  \bibinfo{volume}{17}.
\bibitem[{Lee et~al.(2019)Lee, Nakamura, Narayanan, Brown, Arnold, Initiative
  et~al.}]{lee2019estimating}
\bibinfo{author}{Lee, H.}, \bibinfo{author}{Nakamura, K.},
  \bibinfo{author}{Narayanan, S.}, \bibinfo{author}{Brown, R.A.},
  \bibinfo{author}{Arnold, D.L.}, \bibinfo{author}{Initiative, A.D.N.}, et~al.,
  \bibinfo{year}{2019}.
\newblock \bibinfo{title}{Estimating and accounting for the effect of mri
  scanner changes on longitudinal whole-brain volume change measurements}.
\newblock \bibinfo{journal}{Neuroimage} \bibinfo{volume}{184},
  \bibinfo{pages}{555--565}.
\bibitem[{Lewis and Fox(2004)}]{lewis2004correction}
\bibinfo{author}{Lewis, E.B.}, \bibinfo{author}{Fox, N.C.},
  \bibinfo{year}{2004}.
\newblock \bibinfo{title}{Correction of differential intensity inhomogeneity in
  longitudinal mr images}.
\newblock \bibinfo{journal}{Neuroimage} \bibinfo{volume}{23},
  \bibinfo{pages}{75--83}.
\bibitem[{Li et~al.(2016)Li, Morgan, Ashburner, Smith and Rorden}]{li2016first}
\bibinfo{author}{Li, X.}, \bibinfo{author}{Morgan, P.S.},
  \bibinfo{author}{Ashburner, J.}, \bibinfo{author}{Smith, J.},
  \bibinfo{author}{Rorden, C.}, \bibinfo{year}{2016}.
\newblock \bibinfo{title}{The first step for neuroimaging data analysis: Dicom
  to nifti conversion}.
\newblock \bibinfo{journal}{Journal of neuroscience methods}
  \bibinfo{volume}{264}, \bibinfo{pages}{47--56}.
\bibitem[{Liu et~al.(2022)Liu, Cabezas, Zhan, Wang, Ly, Kyle, Beadnall,
  Butzkueven, Van Der~Walt, Gresle et~al.}]{liu2022dams}
\bibinfo{author}{Liu, D.}, \bibinfo{author}{Cabezas, M.},
  \bibinfo{author}{Zhan, G.}, \bibinfo{author}{Wang, D.}, \bibinfo{author}{Ly,
  L.}, \bibinfo{author}{Kyle, K.}, \bibinfo{author}{Beadnall, H.},
  \bibinfo{author}{Butzkueven, H.}, \bibinfo{author}{Van Der~Walt, A.},
  \bibinfo{author}{Gresle, M.}, et~al., \bibinfo{year}{2022}.
\newblock \bibinfo{title}{Dams-net: A domain adaptive lesion segmentation
  framework in patients with multiple sclerosis from multiple imaging centers
  (p18-4.001)}.
\bibitem[{Liu et~al.(2021)Liu, Maiti, Thomopoulos, Zhu, Chai, Kim and
  Jahanshad}]{liu2021style}
\bibinfo{author}{Liu, M.}, \bibinfo{author}{Maiti, P.},
  \bibinfo{author}{Thomopoulos, S.}, \bibinfo{author}{Zhu, A.},
  \bibinfo{author}{Chai, Y.}, \bibinfo{author}{Kim, H.},
  \bibinfo{author}{Jahanshad, N.}, \bibinfo{year}{2021}.
\newblock \bibinfo{title}{Style transfer using generative adversarial networks
  for multi-site mri harmonization}, in: \bibinfo{booktitle}{International
  Conference on Medical Image Computing and Computer-Assisted Intervention},
  \bibinfo{organization}{Springer}. pp. \bibinfo{pages}{313--322}.
\bibitem[{Lowekamp et~al.(2013)Lowekamp, Chen, Ib{\'a}{\~n}ez and
  Blezek}]{lowekamp2013design}
\bibinfo{author}{Lowekamp, B.C.}, \bibinfo{author}{Chen, D.T.},
  \bibinfo{author}{Ib{\'a}{\~n}ez, L.}, \bibinfo{author}{Blezek, D.},
  \bibinfo{year}{2013}.
\newblock \bibinfo{title}{The design of simpleitk}.
\newblock \bibinfo{journal}{Frontiers in neuroinformatics} \bibinfo{volume}{7},
  \bibinfo{pages}{45}.
\bibitem[{Maclaren et~al.(2014)Maclaren, Han, Vos, Fischbein and
  Bammer}]{maclaren2014reliability}
\bibinfo{author}{Maclaren, J.}, \bibinfo{author}{Han, Z.},
  \bibinfo{author}{Vos, S.B.}, \bibinfo{author}{Fischbein, N.},
  \bibinfo{author}{Bammer, R.}, \bibinfo{year}{2014}.
\newblock \bibinfo{title}{Reliability of brain volume measurements: a
  test-retest dataset}.
\newblock \bibinfo{journal}{Scientific data} \bibinfo{volume}{1},
  \bibinfo{pages}{1--9}.
\bibitem[{Medawar et~al.(2021)Medawar, Thieleking, Manuilova, Paerisch,
  Villringer, Witte and Beyer}]{medawar2021estimating}
\bibinfo{author}{Medawar, E.}, \bibinfo{author}{Thieleking, R.},
  \bibinfo{author}{Manuilova, I.}, \bibinfo{author}{Paerisch, M.},
  \bibinfo{author}{Villringer, A.}, \bibinfo{author}{Witte, A.V.},
  \bibinfo{author}{Beyer, F.}, \bibinfo{year}{2021}.
\newblock \bibinfo{title}{Estimating the effect of a scanner upgrade on
  measures of grey matter structure for longitudinal designs}.
\newblock \bibinfo{journal}{PloS one} \bibinfo{volume}{16},
  \bibinfo{pages}{e0239021}.
\bibitem[{Miyato et~al.(2018)Miyato, Maeda, Koyama and
  Ishii}]{miyato2018virtual}
\bibinfo{author}{Miyato, T.}, \bibinfo{author}{Maeda, S.i.},
  \bibinfo{author}{Koyama, M.}, \bibinfo{author}{Ishii, S.},
  \bibinfo{year}{2018}.
\newblock \bibinfo{title}{Virtual adversarial training: a regularization method
  for supervised and semi-supervised learning}.
\newblock \bibinfo{journal}{IEEE transactions on pattern analysis and machine
  intelligence} \bibinfo{volume}{41}, \bibinfo{pages}{1979--1993}.
\bibitem[{Popescu et~al.(2013)Popescu, Agosta, Hulst, Sluimer, Knol, Sormani,
  Enzinger, Ropele, Alonso, Sastre-Garriga et~al.}]{popescu2013brain}
\bibinfo{author}{Popescu, V.}, \bibinfo{author}{Agosta, F.},
  \bibinfo{author}{Hulst, H.E.}, \bibinfo{author}{Sluimer, I.C.},
  \bibinfo{author}{Knol, D.L.}, \bibinfo{author}{Sormani, M.P.},
  \bibinfo{author}{Enzinger, C.}, \bibinfo{author}{Ropele, S.},
  \bibinfo{author}{Alonso, J.}, \bibinfo{author}{Sastre-Garriga, J.}, et~al.,
  \bibinfo{year}{2013}.
\newblock \bibinfo{title}{Brain atrophy and lesion load predict long term
  disability in multiple sclerosis}.
\newblock \bibinfo{journal}{Journal of Neurology, Neurosurgery \& Psychiatry}
  \bibinfo{volume}{84}, \bibinfo{pages}{1082--1091}.
\bibitem[{Prados et~al.(2015)Prados, Cardoso, Leung, Cash, Modat, Fox,
  Wheeler-Kingshott, Ourselin, Initiative et~al.}]{prados2015measuring}
\bibinfo{author}{Prados, F.}, \bibinfo{author}{Cardoso, M.J.},
  \bibinfo{author}{Leung, K.K.}, \bibinfo{author}{Cash, D.M.},
  \bibinfo{author}{Modat, M.}, \bibinfo{author}{Fox, N.C.},
  \bibinfo{author}{Wheeler-Kingshott, C.A.}, \bibinfo{author}{Ourselin, S.},
  \bibinfo{author}{Initiative, A.D.N.}, et~al., \bibinfo{year}{2015}.
\newblock \bibinfo{title}{Measuring brain atrophy with a generalized
  formulation of the boundary shift integral}.
\newblock \bibinfo{journal}{Neurobiology of aging} \bibinfo{volume}{36},
  \bibinfo{pages}{S81--S90}.
\bibitem[{Preboske et~al.(2006)Preboske, Gunter, Ward and
  Jack~Jr}]{preboske2006common}
\bibinfo{author}{Preboske, G.M.}, \bibinfo{author}{Gunter, J.L.},
  \bibinfo{author}{Ward, C.P.}, \bibinfo{author}{Jack~Jr, C.R.},
  \bibinfo{year}{2006}.
\newblock \bibinfo{title}{Common mri acquisition non-idealities significantly
  impact the output of the boundary shift integral method of measuring brain
  atrophy on serial mri}.
\newblock \bibinfo{journal}{Neuroimage} \bibinfo{volume}{30},
  \bibinfo{pages}{1196--1202}.
\bibitem[{Reuter et~al.(2012)Reuter, Schmansky, Rosas and
  Fischl}]{reuter2012within}
\bibinfo{author}{Reuter, M.}, \bibinfo{author}{Schmansky, N.J.},
  \bibinfo{author}{Rosas, H.D.}, \bibinfo{author}{Fischl, B.},
  \bibinfo{year}{2012}.
\newblock \bibinfo{title}{Within-subject template estimation for unbiased
  longitudinal image analysis}.
\newblock \bibinfo{journal}{Neuroimage} \bibinfo{volume}{61},
  \bibinfo{pages}{1402--1418}.
\bibitem[{Rudick et~al.(1999)Rudick, Fisher, Lee, Simon, Jacobs, Group
  et~al.}]{rudick1999use}
\bibinfo{author}{Rudick, R.}, \bibinfo{author}{Fisher, E.},
  \bibinfo{author}{Lee, J.C.}, \bibinfo{author}{Simon, J.},
  \bibinfo{author}{Jacobs, L.}, \bibinfo{author}{Group, M.S.C.R.}, et~al.,
  \bibinfo{year}{1999}.
\newblock \bibinfo{title}{Use of the brain parenchymal fraction to measure
  whole brain atrophy in relapsing-remitting ms}.
\newblock \bibinfo{journal}{Neurology} \bibinfo{volume}{53},
  \bibinfo{pages}{1698--1698}.
\bibitem[{Rudick et~al.(2000)Rudick, Fisher, Lee, Duda and
  Simon}]{rudick2000brain}
\bibinfo{author}{Rudick, R.A.}, \bibinfo{author}{Fisher, E.},
  \bibinfo{author}{Lee, J.C.}, \bibinfo{author}{Duda, J.T.},
  \bibinfo{author}{Simon, J.}, \bibinfo{year}{2000}.
\newblock \bibinfo{title}{Brain atrophy in relapsing multiple sclerosis:
  relationship to relapses, edss, and treatment with interferon $\beta$-1a}.
\newblock \bibinfo{journal}{Multiple Sclerosis Journal} \bibinfo{volume}{6},
  \bibinfo{pages}{365--372}.
\bibitem[{Sajjadi et~al.(2016)Sajjadi, Javanmardi and
  Tasdizen}]{sajjadi2016regularization}
\bibinfo{author}{Sajjadi, M.}, \bibinfo{author}{Javanmardi, M.},
  \bibinfo{author}{Tasdizen, T.}, \bibinfo{year}{2016}.
\newblock \bibinfo{title}{Regularization with stochastic transformations and
  perturbations for deep semi-supervised learning}.
\newblock \bibinfo{journal}{Advances in neural information processing systems}
  \bibinfo{volume}{29}.
\bibitem[{Shorten and Khoshgoftaar(2019)}]{shorten2019survey}
\bibinfo{author}{Shorten, C.}, \bibinfo{author}{Khoshgoftaar, T.M.},
  \bibinfo{year}{2019}.
\newblock \bibinfo{title}{A survey on image data augmentation for deep
  learning}.
\newblock \bibinfo{journal}{Journal of big data} \bibinfo{volume}{6},
  \bibinfo{pages}{1--48}.
\bibitem[{Sinnecker et~al.(2022)Sinnecker, Sch{\"a}delin, Benkert, Ruberte,
  Amann, Lieb, Naegelin, M{\"u}ller, Kuhle, Derfuss
  et~al.}]{sinnecker2022brain}
\bibinfo{author}{Sinnecker, T.}, \bibinfo{author}{Sch{\"a}delin, S.},
  \bibinfo{author}{Benkert, P.}, \bibinfo{author}{Ruberte, E.},
  \bibinfo{author}{Amann, M.}, \bibinfo{author}{Lieb, J.M.},
  \bibinfo{author}{Naegelin, Y.}, \bibinfo{author}{M{\"u}ller, J.},
  \bibinfo{author}{Kuhle, J.}, \bibinfo{author}{Derfuss, T.}, et~al.,
  \bibinfo{year}{2022}.
\newblock \bibinfo{title}{Brain atrophy measurement over a mri scanner change
  in multiple sclerosis}.
\newblock \bibinfo{journal}{NeuroImage: Clinical} \bibinfo{volume}{36},
  \bibinfo{pages}{103148}.
\bibitem[{Smeets et~al.(2016)Smeets, Ribbens, Sima, Cambron, Horakova, Jain,
  Maertens, Van~Vlierberghe, Terzopoulos, Van~Binst
  et~al.}]{smeets2016reliable}
\bibinfo{author}{Smeets, D.}, \bibinfo{author}{Ribbens, A.},
  \bibinfo{author}{Sima, D.M.}, \bibinfo{author}{Cambron, M.},
  \bibinfo{author}{Horakova, D.}, \bibinfo{author}{Jain, S.},
  \bibinfo{author}{Maertens, A.}, \bibinfo{author}{Van~Vlierberghe, E.},
  \bibinfo{author}{Terzopoulos, V.}, \bibinfo{author}{Van~Binst, A.M.}, et~al.,
  \bibinfo{year}{2016}.
\newblock \bibinfo{title}{Reliable measurements of brain atrophy in individual
  patients with multiple sclerosis}.
\newblock \bibinfo{journal}{Brain and behavior} \bibinfo{volume}{6},
  \bibinfo{pages}{e00518}.
\bibitem[{Smith et~al.(2002)Smith, Zhang, Jenkinson, Chen, Matthews, Federico
  and De~Stefano}]{smith2002accurate}
\bibinfo{author}{Smith, S.M.}, \bibinfo{author}{Zhang, Y.},
  \bibinfo{author}{Jenkinson, M.}, \bibinfo{author}{Chen, J.},
  \bibinfo{author}{Matthews, P.M.}, \bibinfo{author}{Federico, A.},
  \bibinfo{author}{De~Stefano, N.}, \bibinfo{year}{2002}.
\newblock \bibinfo{title}{Accurate, robust, and automated longitudinal and
  cross-sectional brain change analysis}.
\newblock \bibinfo{journal}{Neuroimage} \bibinfo{volume}{17},
  \bibinfo{pages}{479--489}.
\bibitem[{Takao et~al.(2010)Takao, Abe, Hayashi, Kabasawa and
  Ohtomo}]{takao2010effects}
\bibinfo{author}{Takao, H.}, \bibinfo{author}{Abe, O.},
  \bibinfo{author}{Hayashi, N.}, \bibinfo{author}{Kabasawa, H.},
  \bibinfo{author}{Ohtomo, K.}, \bibinfo{year}{2010}.
\newblock \bibinfo{title}{Effects of gradient non-linearity correction and
  intensity non-uniformity correction in longitudinal studies using structural
  image evaluation using normalization of atrophy (siena)}.
\newblock \bibinfo{journal}{Journal of Magnetic Resonance Imaging}
  \bibinfo{volume}{32}, \bibinfo{pages}{489--492}.
\bibitem[{Tedeschi et~al.(2005)Tedeschi, Lavorgna, Russo, Prinster, Dinacci,
  Savettieri, Quattrone, Livrea, Messina, Reggio et~al.}]{tedeschi2005brain}
\bibinfo{author}{Tedeschi, G.}, \bibinfo{author}{Lavorgna, L.},
  \bibinfo{author}{Russo, P.}, \bibinfo{author}{Prinster, A.},
  \bibinfo{author}{Dinacci, D.}, \bibinfo{author}{Savettieri, G.},
  \bibinfo{author}{Quattrone, A.}, \bibinfo{author}{Livrea, P.},
  \bibinfo{author}{Messina, C.}, \bibinfo{author}{Reggio, A.}, et~al.,
  \bibinfo{year}{2005}.
\newblock \bibinfo{title}{Brain atrophy and lesion load in a large population
  of patients with multiple sclerosis}.
\newblock \bibinfo{journal}{Neurology} \bibinfo{volume}{65},
  \bibinfo{pages}{280--285}.
\bibitem[{Thanellas and Pollari(2010)}]{thanellas2010sensitivity}
\bibinfo{author}{Thanellas, A.K.}, \bibinfo{author}{Pollari, M.},
  \bibinfo{year}{2010}.
\newblock \bibinfo{title}{Sensitivity of volumetric brain analysis to
  systematic and random errors}, in: \bibinfo{booktitle}{2010 IEEE 23rd
  International Symposium on Computer-Based Medical Systems (CBMS)},
  \bibinfo{organization}{IEEE}. pp. \bibinfo{pages}{238--242}.
\bibitem[{Van~Leemput et~al.(1999)Van~Leemput, Maes, Vandermeulen and
  Suetens}]{van1999automated}
\bibinfo{author}{Van~Leemput, K.}, \bibinfo{author}{Maes, F.},
  \bibinfo{author}{Vandermeulen, D.}, \bibinfo{author}{Suetens, P.},
  \bibinfo{year}{1999}.
\newblock \bibinfo{title}{Automated model-based tissue classification of mr
  images of the brain}.
\newblock \bibinfo{journal}{IEEE transactions on medical imaging}
  \bibinfo{volume}{18}, \bibinfo{pages}{897--908}.
\bibitem[{Vemuri et~al.(2005)Vemuri, Kholmovski, Parker and
  Chapman}]{vemuri2005coil}
\bibinfo{author}{Vemuri, P.}, \bibinfo{author}{Kholmovski, E.G.},
  \bibinfo{author}{Parker, D.L.}, \bibinfo{author}{Chapman, B.E.},
  \bibinfo{year}{2005}.
\newblock \bibinfo{title}{Coil sensitivity estimation for optimal snr
  reconstruction and intensity inhomogeneity correction in phased array mr
  imaging}, in: \bibinfo{booktitle}{Biennial international conference on
  information processing in medical imaging}, \bibinfo{organization}{Springer}.
  pp. \bibinfo{pages}{603--614}.
\bibitem[{Vovk et~al.(2004)Vovk, Pernu{\v{s}} and Likar}]{vovk2004mri}
\bibinfo{author}{Vovk, U.}, \bibinfo{author}{Pernu{\v{s}}, F.},
  \bibinfo{author}{Likar, B.}, \bibinfo{year}{2004}.
\newblock \bibinfo{title}{Mri intensity inhomogeneity correction by combining
  intensity and spatial information}.
\newblock \bibinfo{journal}{Physics in Medicine \& Biology}
  \bibinfo{volume}{49}, \bibinfo{pages}{4119}.
\bibitem[{Vrenken et~al.(2013)Vrenken, Jenkinson, Horsfield, Battaglini,
  Van~Schijndel, Rostrup, Geurts, Fisher, Zijdenbos, Ashburner
  et~al.}]{vrenken2013recommendations}
\bibinfo{author}{Vrenken, H.}, \bibinfo{author}{Jenkinson, M.},
  \bibinfo{author}{Horsfield, M.}, \bibinfo{author}{Battaglini, M.},
  \bibinfo{author}{Van~Schijndel, R.}, \bibinfo{author}{Rostrup, E.},
  \bibinfo{author}{Geurts, J.}, \bibinfo{author}{Fisher, E.},
  \bibinfo{author}{Zijdenbos, A.}, \bibinfo{author}{Ashburner, J.}, et~al.,
  \bibinfo{year}{2013}.
\newblock \bibinfo{title}{Recommendations to improve imaging and analysis of
  brain lesion load and atrophy in longitudinal studies of multiple sclerosis}.
\newblock \bibinfo{journal}{Journal of neurology} \bibinfo{volume}{260},
  \bibinfo{pages}{2458--2471}.
\bibitem[{Wu and He(2018)}]{wu2018group}
\bibinfo{author}{Wu, Y.}, \bibinfo{author}{He, K.}, \bibinfo{year}{2018}.
\newblock \bibinfo{title}{Group normalization}, in:
  \bibinfo{booktitle}{Proceedings of the European conference on computer vision
  (ECCV)}, pp. \bibinfo{pages}{3--19}.
\bibitem[{Zhang et~al.(2017a)Zhang, Cisse, Dauphin and
  Lopez-Paz}]{zhang2017mixup}
\bibinfo{author}{Zhang, H.}, \bibinfo{author}{Cisse, M.},
  \bibinfo{author}{Dauphin, Y.N.}, \bibinfo{author}{Lopez-Paz, D.},
  \bibinfo{year}{2017}a.
\newblock \bibinfo{title}{mixup: Beyond empirical risk minimization}.
\newblock \bibinfo{journal}{arXiv preprint arXiv:1710.09412} .
\bibitem[{Zhang et~al.(2017b)Zhang, Hou, Wang and Song}]{zhang2017regularizing}
\bibinfo{author}{Zhang, S.}, \bibinfo{author}{Hou, Y.}, \bibinfo{author}{Wang,
  B.}, \bibinfo{author}{Song, D.}, \bibinfo{year}{2017}b.
\newblock \bibinfo{title}{Regularizing neural networks via retaining confident
  connections}.
\newblock \bibinfo{journal}{Entropy} \bibinfo{volume}{19},
  \bibinfo{pages}{313}.
\bibitem[{Zhang et~al.(2001)Zhang, Brady and Smith}]{zhang2001segmentation}
\bibinfo{author}{Zhang, Y.}, \bibinfo{author}{Brady, M.},
  \bibinfo{author}{Smith, S.}, \bibinfo{year}{2001}.
\newblock \bibinfo{title}{Segmentation of brain mr images through a hidden
  markov random field model and the expectation-maximization algorithm}.
\newblock \bibinfo{journal}{IEEE transactions on medical imaging}
  \bibinfo{volume}{20}, \bibinfo{pages}{45--57}.
\bibitem[{Zhou et~al.(2022)Zhou, Liu, Qiao, Xiang and Loy}]{zhou2022domain}
\bibinfo{author}{Zhou, K.}, \bibinfo{author}{Liu, Z.}, \bibinfo{author}{Qiao,
  Y.}, \bibinfo{author}{Xiang, T.}, \bibinfo{author}{Loy, C.C.},
  \bibinfo{year}{2022}.
\newblock \bibinfo{title}{Domain generalization: A survey}.
\newblock \bibinfo{journal}{IEEE Transactions on Pattern Analysis and Machine
  Intelligence} .

\end{thebibliography}

\bio{}
\endbio

\endbio

\end{document}